\documentclass[
 reprint,floatfix,
 amsmath,amssymb,
 aps,prb,superscriptaddress,longbibliography,nofootinbib
]{revtex4-2}

\usepackage{amsthm,bm,mathtools}
\usepackage{tikz,graphicx,dcolumn}
\usepackage{standalone}
\usepackage{hyperref}
\usepackage{float}
\usepackage{ulem}
\usepackage{units}

\usepackage[utf8]{inputenc}
\newcommand{\angstrom}{\text{\normalfont\AA}}
\usepackage[english]{babel}

\newcommand{\beginsupplement}{%
        \setcounter{table}{0}
        \renewcommand{\thetable}{S\arabic{table}}%
        \setcounter{figure}{0}
        \renewcommand{\thefigure}{S\arabic{figure}}%
     }

\DeclarePairedDelimiter\norm{\lVert}{\rVert}%
\let\oldnorm\norm
\def\norm{\@ifstar{\oldnorm}{\oldnorm*}}
\makeatother

\begin{document}

\preprint{APS/123-QED}

\title{A feasibility analysis towards the simulation of hysteresis with spin-lattice dynamics}

\author{G. dos Santos}
\affiliation{CONICET, Mendoza, 5500, Argentina}
\affiliation{Facultad de Ingeniería, Universidad de Mendoza, Mendoza, 5500 Argentina}
\email{gonzalodossantos@gmail.com}

\author{F. Rom\'{a}}
\affiliation{Departamento de F\'{i}sica, Universidad Nacional de San Luis, Instituto de F\'{i}sica Aplicada (INFAP), Consejo Nacional de Investigaciones
Cient\'{i}ficas y T\'{e}cnicas (CONICET), Chacabuco 917, D5700BWS San Luis, Argentina}

\author{J. Tranchida}
\affiliation{CEA, DES, IRESNE, DEC, SESC, LM2C, F-13108 Saint-Paul-Lez-Durance, France.}

\author{S. Castedo}
\affiliation{
Department of Physics and
Astronomy, School of Natural Sciences, The University of Manchester, Manchester M13
9PL, United Kingdom}

\author{L. F. Cugliandolo}
\affiliation{Sorbonne Université, CNRS UMR 7589, Laboratoire de Physique Théorique et Hautes Energies, 4 Place Jussieu, 75252 Paris  Cedex 05, France}
\affiliation{Institut Universitaire de France, 1 rue Descartes, 75231 Paris Cedex 05, France}

\author{E. M. Bringa}
\affiliation{CONICET, Mendoza, 5500, Argentina}
\affiliation{Facultad de Ingeniería, Universidad de Mendoza, Mendoza, 5500 Argentina}
\affiliation{Centro de Nanotecnolog\'{i}a Aplicada, Facultad de Ciencias, Universidad Mayor, Santiago, Chile 8580745}

\date{\today}

\begin{abstract}
We use spin-lattice dynamics simulations to study the possibility of modeling 
the magnetic hysteresis behavior of a ferromagnetic material. The temporal evolution of the magnetic and mechanical degrees of freedom is obtained through a set of two coupled Langevin equations. Hysteresis loops are calculated for different angles between the external field and the magnetocrystalline anisotropy axes. The influence of several relevant parameters is studied, including the field frequency, magnetic damping, magnetic anisotropy (magnitude and type), magnetic exchange, and system size. The role played by a moving lattice is also discussed.  For a perfect bulk ferromagnetic system we find that, at low temperatures, the exchange and lattice dynamics barely affect the loops, 
while the field frequency and magnetic damping have a large effect on it. The influence of the anisotropy magnitude and symmetry are found to follow the expected behavior. We show that a careful choice of simulation parameters allows for an excellent agreement between the spin-lattice dynamics measurements and the paradigmatic Stoner-Wohlfarth model. \textcolor{black}{Furthermore, we extend this analysis to intermediate and high temperatures for the perfect bulk system and for spherical nanoparticles, with and without defects, reaching values close to the Curie temperature. In this temperature range, we find that lattice dynamics has a greater role on the magnetic behavior, especially in the evolution of the defective samples}. The present study opens the possibility for more accurate inclusion of lattice defects and thermal effects in hysteresis simulations.
\end{abstract}
\maketitle

\noindent
{\bf Keywords} Hysteresis, spin-lattice dynamics, ferromagnetic, Fe

\section{\label{sec:introduction} Introduction}

For basic research and applications, magnetic studies are of fundamental relevance as they are an important tool for revealing otherwise hidden structural, thermodynamic and physicochemical properties of a material. Hysteresis loops are usually measured to characterize the dynamic behavior of bulk samples that exhibit microstructures like grain boundaries, defects, etc, and multiple domains \cite{Bertotti1998Hysteresis,Mayergoyz2003Hysteresis}. Furthermore, such measurements are also performed for low-dimensional magnets such as magnetic nanoparticles (NPs) or thin films \cite{cullity2011introduction}. 

The power generated by a sample subject to an alternating magnetic field can be directly determined from the shape of a hysteresis loop. In fact, the Specific Absorption Rate (SAR), defined as the absorbed energy per unit of mass, is proportional to the area of this curve \cite{mehdaoui2011optimal}. In magnetic hyperthermia \cite{shaterabadi2018physics, pankhurst2003applications, hedayatnasab2017review}, for instance, SAR provides the heating efficiency of a specific type of magnetic nanoparticle. \textcolor{black}{Furthermore, the magnetization of nanoparticles is of great interest for many other technological applications \cite{barman2020magnetization, ma2021_NP-review}. Magnetization in small NPs is expected to flip due to thermal fluctuations during the time scale of a few seconds in typical experiments, giving rise to superparamagnetic behavior below some critical size, which for Fe is close to 20 nm. However, hysteresis loops with ferromagnetic blocked behavior  have been observed for relatively simple Fe NPs, indicating complex magnetic behavior, which is not fully understood \cite{ibusuki2001magnetic, balan2014direct, kleibert2017direct}. Some of this behavior might be related to lattice defects \cite{balan2014direct, kleibert2017direct}. It has been emphasized that defect engineering will allow novel future technological applications of magnetic nanoparticles \cite{batlle2022magnetic, lak2021embracing}, pointing out the need to model defective magnetic nanosystems.}

Different classical models are used to describe magnetic hysteresis  \cite{Bertotti1998Hysteresis,liorzou2000macroscopic}.  Possibly, the most popular one is that of Stoner-Wohlfarth (SW) \cite{stoner1948mechanism,tannous2008stoner}. This model is relatively simple to evaluate numerically, allowing easy comparison with experimental results. However, the SW model assumes an important number of approximations. Among them, it is considered that the material behaves as a single magnetic domain. This is known as the macrospin approximation, which implies that below a certain length scale, typically on the order of a few nanometers, the internal structure of the material is neglected. In the SW model the Hamiltonian is the sum of a Zeeman term and an uniaxial anisotropy term. Hysteretic behavior arises as a consequence of a simple zero-temperature dynamic protocol: as the external field varies in time, the magnetization can either change continuously while the system remains at an energy minimum, or it can jump abruptly to another value when that minimum becomes unstable. 
Usually the SW model is employed to compare with experiments for a large collection of randomly oriented NP, and obtain relevant physical quantities \cite{ibusuki2001magnetic}. In addition, although in the original SW model thermal fluctuations were not considered, further approaches have attempted to incorporate temperature effects \cite{lanci2003introduction}. 

Fortunately, micromagnetic simulations allow one to overcome many of these limitations \cite{Bertotti-Mayergoyz-Serpico2008,behbahani2020coarse}. Thus, it is possible to study a system made up of a large number of interacting macrospins with a Hamiltonian including any complex energetic term. In particular, the equilibrium or nonequilibrium magnetization dynamics of such a model can be calculated by numerically solving either the stochastic Landau-Lifshitz-Gilbert (sLLG) or 
the stochastic Landau-Lifshitz (sLL) phenomenological equations \cite{Brown1963-sLLG,garcia1998langevin}. From {\it first principles} it can be proved that this same theoretical framework is valid to carry out atomistic spin dynamic (ASD) simulations \cite{evans2014atomistic,Atomistic-Spin-Dynamics-book}. This simulation method has been successfully applied to the modeling of hysteresis loops \cite{bottcher2011atomistic,alzate2019optimal,westmoreland2020atomistic,jenkins2021atomistic}. Another alternative to include temperature effects in hysteresis loops is to use Metropolis Monte Carlo simulations for a fixed lattice \cite{tajiri2018effect,essajai2019shape}. 

More recently, there has been increased interest in studying more realistic models that take into account both spin and lattice degrees of freedom, and the coupling between them. Most of these methods are based on coupled Spin Dynamics (SD) and Molecular Dynamics (MD) simulations \cite{shimada2015magnetic,ma2016spilady,wu2018magnon,dosSantos2020,dosSantos2021}. This is commonly known as spin-lattice dynamics (SLD) simulations. Other first-principle based methods combine atomistic spin dynamics with ab initio molecular dynamics (ASD-AIMD) \cite{stockem2018anomalous}, but this is limited to systems with a few atoms due to the high computational cost of calculating interatomic forces via ab initio methods.

  SLD simulations are typically based on a Langevin approach that allows one to describe the evolution of both lattice and spin degrees of freedom \cite{tranchida2018massively}. Although this method is very powerful, its implementation can be computationally demanding and, as the complexity of a system increases, the number of parameters that characterize it also increases. Therefore, the practical use of this numerical scheme for calculation of hysteresis loops requires a deep exploration. In this work, we use SLD calculations with sLL spin dynamics to simulate the hysteretic behavior of a ferromagnetic material. We use physical parameters that are typical for bulk bcc iron. We focus on determining optimal simulation parameters that allow one to obtain reliable results at low temperatures.  Also, we explore how the hysteresis loops depend on the different physical properties that characterize these magnetic systems: anisotropy, exchange, damping, and lattice vibrations. The paradigmatic SW model is taken as a reference to achieve this goal. Additional simulations are carried out to analyze the importance of SLD simulation at higher temperatures. \textcolor{black}{Finally, we present NP hysteresis loops simulations, including defects like the ones found in experiments.}
  
We explore the possibility of using a high damping parameter in order to reduce the relaxation time and speed up the calculations. The use of strong damping led us to revise the definition of the noise-noise correlations in the sLL equations and find the correct parameter dependence that lets the systems thermalize with their equilibrium environments at long times. In this work, we perform SLD simulations using the SPIN package of the software LAMMPS \cite{thompson2022lammps, tranchida2018massively, LAMMPS-spin}.In order to enable its use in this study with relatively large damping, the software was modified accordingly. 
 
 The paper is organized as follows. In Sec. \ref{sec:methods}, the theoretical models used and the simulation details are presented. In Sec. \ref{sec:results}, the main results of this study are shown. A summary and the conclusions of the work are drawn in Sec. \ref{sec:conclusions}. In order to simplify several parametric studies, as a first approach we run spin dynamics simulations with the positions of the atoms fixed at their ideal equilibrium values. The influence of the spin-lattice coupling is explored and discussed in the last part of Sec. \ref{sec:results}.
Finally, in an Appendix we derive the Fokker-Planck equation associated to the SLD dynamics and 
from it, we fix the noise-noise correlations that ensure the asymptotic approach to equilibrium.
We confirmed, with simulations not shown here, that these do indeed take the systems to equilibrium even at large values of the damping factor, very convenient to reduce the simulation time.

\section{\label{sec:methods} Methods}
In Sec.~\ref{sec:framework} we describe the main characteristics of the SLD model from LAMMPS\cite{thompson2022lammps, tranchida2018massively, LAMMPS-spin}, and in Sec.~\ref{sec:simulation_details} we give some simulation details. 

\subsection{\label{sec:framework} SLD Model}

Let us consider an ensemble of $N$ atoms 
each endowed with a classical magnetic moment.
The Hamiltonian of the model is
\begin{equation}
  \mathcal{H} =
  \sum_{i=1}^{N} \frac{\left| \bm{p}_i \right|^2}{2m_i} +
  \sum_{i,j,i\neq j}^{N} V(r_{ij}) + \mathcal{H}_{\rm mag}.
  \label{hamiltonian}
\end{equation}
The first term in Eq.~(\ref{hamiltonian}) accounts for 
the kinetic energy of the $N$ atoms
where $\bm{p}_i$ and $m_i$ represent, respectively, 
the linear momentum and mass of the $i$-th atom ($m_i=55.845$ u for iron).  
The second term is the interatomic potential describing 
the interactions between pairs of atoms 
at positions $\bm{r}_i$ and $\bm{r}_j$ separated by a distance $r_{ij} \equiv |\bm{r}_i-\bm{r}_j|$. 
As usual in molecular dynamics simulations of metals, 
we use a classical embedded atom model (EAM) potential
which describes well a broad spectrum of iron properties \cite{chamati2006embedded, chamatiPot-NIST}. The interatomic cutoff distance for this potential was set to $0.57$ nm. 

The coupling among the spin and lattice degrees of freedom
is provided through the last term in Eq.~(\ref{hamiltonian}), 
a magnetic Hamiltonian defined as 
\begin{equation}
  \mathcal{H}_{\rm mag} = 
 - \mu_0 \mu \bm{H} \cdot \sum_{i=0}^{N} \bm{s}_i
 - \sum_{i,j,i\neq j}^{N} J(r_{ij}) \bm{s}_i\cdot \bm{s}_j 
 + \mathcal{H}_{\rm ani}.
  \label{hamiltonian_mag}
\end{equation}
Here, $\bm{s}_i$ is a classical unitary vector representing the spin of the $i$-th atom. The first term in Eq.~(\ref{hamiltonian_mag}) is the Zeeman energy, 
the interaction of each spin with an external uniform magnetic field $\bm{H}$, where
$\mu=2.2 \mu_B$ is the atomic magnetic moment for iron
($\mu_B$ is the Bohr magneton) and $\mu_0$ is the vacuum permeability constant.  
Note that the Zeeman energy within LAMMPS is defined slightly differently than in Eq.~(\ref{hamiltonian_mag}), and a re-scaling had to be applied to recover our formulation. 
The second term is just a Heisenberg Hamiltonian 
describing the interaction between spins, where $J(r_{ij})$ is an interatomic distance-dependent exchange coupling which is defined as the following Bethe-Slater curve \cite{kaneyoshi1992introduction, yosida1996theory},
\begin{eqnarray}
  J(r_{ij}) &=& 
  4 \alpha \left(\frac{r_{ij}}{\delta} \right)^2 \left[1-\gamma 
  \left(\frac{r_{ij}}{\delta} \right)^2 \right] e^{-\left(\frac{r_{ij}}{\delta} \right)^2}
  \nonumber\\
  &&
  \; \times \; \Theta(R_c - r_{ij}),
  \label{exchange_dep}
\end{eqnarray}
being $\Theta(R_c - r_{ij})$ the Heaviside step function and $R_c$ the cutoff distance. The coefficients in Eq.~(\ref{exchange_dep}) can be fitted with ab-initio or experimental data. In the present work we fit Eq.~(\ref{exchange_dep}) to data by Ma {\it et al.}~\cite{ma2008large}, as it was already done in previous works \cite{dosSantos2020, meyer2022influence}. \textcolor{black}{Specifically, we use the following values: $\alpha = 96.0$ meV, $\gamma = 0.20$, $\delta = 0.154$ nm.} In addition, we choose $R_c=0.35$~nm. 

Finally, the last term in Eq.~(\ref{hamiltonian_mag}) is responsible 
for computing the magneto-crystalline anisotropy. In this work we consider one of two, either uniaxial $\mathcal{H}_{\rm uni}$ or cubic $\mathcal{H}_{\rm cub}$ anisotropy. The corresponding expressions are given by
\begin{equation}
    \mathcal{H}_{\rm uni} = - K_1 \sum_{i=1}^{N} 
    \; (\bm{s}_i \cdot \bm{n})^2
    \label{hamiltonian_uniaxial}
\end{equation}
and
\begin{eqnarray}
        \mathcal{H}_{\rm cub} 
        &=&  
        \sum_{i=1}^{N} \; \Big{\{} K_1 \, [ (\bm{s}_i \cdot \bm{n}_1)^2 (\bm{s}_i \cdot \bm{n}_2)^2 
        \nonumber\\
        && 
        \qquad \qquad
        + (\bm{s}_i \cdot \bm{n}_2)^2 (\bm{s}_i \cdot \bm{n}_3)^2 
        \nonumber\\
        [10pt]
        && 
        \qquad \qquad
        + (\bm{s}_i \cdot \bm{n}_1)^2 (\bm{s}_i \cdot \bm{n}_3)^2 ] 
        \nonumber\\
        [10pt]
        && 
        \quad\quad\;
        - K_2 \; (\bm{s}_i \cdot \bm{n}_1)^2(\bm{s}_i \cdot \bm{n}_2)^2(\bm{s}_i \cdot \bm{n}_3)^2 
        \Big{\}} 
        \; .
    \label{hamiltonian_aniso}
\end{eqnarray}
Here, the unit vectors $\bm{n}_1$, $\bm{n}_2$, and $\bm{n}_3$ 
lie along the three crystallographic directions $[100]$, $[010]$, and $[001]$, respectively. In Eq.~(\ref{hamiltonian_uniaxial}), $\bm{n}$ is also a unit vector that in general could point along any of these axes. 
The first term in the cubic anisotropy energy above is defined with different 
sign within LAMMPS. $K_1$ and $K_2$ are the magneto crystalline anisotropy constants which we set to $K_1=35$ $\mu$eV/atom and  $K_2=3.6$ $\mu$eV/atom
(equivalents to volumetric anisotropies $K_{1V}=470$ kJ/m$^{3}$
and $K_{2V}=46$ kJ/m$^{3}$). In order to make comparisons, we have used the same value of $K_1$ in both anisotropy equations. Since this constant is positive (and also $K_2>0$), then the easy axes of magnetization
in Eqs.~(\ref{hamiltonian_uniaxial}) and (\ref{hamiltonian_aniso})
are given by the unit vector defined above. Note that the values of $K_1$ and $K_2$ are ten times larger than those usually used to model bulk bcc iron \cite{cullity2011introduction}. However, such larger anisotropy magnitude has been considered for Fe nanoparticles \cite{ibusuki2001magnetic, pastor1995magnetic}, and we also consider bulk values for selected runs. 
As we will discuss later, using a large anisotropy in the simulations 
helps to quickly stabilize the magnetization of the system. \textcolor{black}{Increasing anisotropy for numerical reasons has been employed for other magnetic simulations, for instance, to obtain domain walls with smaller widths \cite{chauleau2020electric} or to confine the magnetization dynamics to a plane \cite{frej2023phonon}.}

We note that an improved implementation of the anisotropy would include a better description of the spin-orbit coupling \cite{strungaru2021spin,caturla-dednam2022spin,cooke2023angular} and also a local variation of magnetic moments and couplings, for instance, depending on atomic volume \cite{nieves2021neel}. However, uniaxial and cubic anisotropies are often used to interpret experimental results, for a number of models and simulations \cite{ibusuki2001magnetic, wang2012temperature, goiriena2020disk, carrey2011simple, hergt2004maghemite}. We include a simple description of anisotropy within this spirit and, in our simulations, lattice and spin dynamics are coupled through the distance-dependent exchange function $J(r_{ij})$ and by the spins Langevin thermostat set at the same temperature as the lattice (see below). In addition, to speed up our calculations we have neglected long-range dipolar interactions in the magnetic Hamiltonian. Since we simulate small systems (see below), this approximation should not affect the validity of our results.

The core of this simulation method, lattice and spin coupling, is contained in the following coupled Langevin equations \cite{tranchida2018massively},
\begin{eqnarray}
  \frac{d\bm{r}_i}{dt} &=& \frac{\bm{p}_i}{m_i} \label{EOM-p1} \, , \\
  \frac{d\bm{p}_i}{dt} &=& \sum_{i,j,i\neq j}^{N} \left[  -\frac{dV \left(r_{ij}\right)}{dr_{ij}} + \frac{dJ\left(r_{ij}\right)}{dr_{ij}} \bm{s}_i \cdot \bm{s}_j  \right]\bm{e}_{ij} \nonumber \\ 
  &&- \frac{\gamma_L}{m_i} \bm{p}_i + \bm{\xi}_i   \label{EOM-p2} \, ,  \\ 
  \frac{d\bm{s}_i}{dt} &=& \frac{1}{1+\lambda_s^2} \left[ \left( \bm{\omega_i} + \bm{\zeta}_i \right) \times \bm{s}_i + \lambda_s\bm{s}_i \times \left(\bm{\omega}_i \times \bm{s}_i\right)  \right]   \,  . \label{EOM_spin2}
\end{eqnarray}

 In Eq.~(\ref{EOM-p2}), $\bm{e}_{ij}$ represents a unit vector along the line connecting atoms $i$ and $j$, $\gamma_L$ is the lattice damping coefficient, and $\bm{\xi}(t)$ is a random fluctuating force drawn from a Gaussian distribution with 
\begin{eqnarray}
\langle\bm{\xi}(t)\rangle &=& 0 
\; , 
\nonumber \\
\langle \xi_a(t) \xi_b(t') \rangle &=& 2 D_L \delta_{a b} \delta(t-t')
\; , 
\end{eqnarray}
where the $a$ and $b$ subscripts indicate Cartesian vector components, and the amplitude of the noise is
\begin{equation}
D_L=\gamma_L k_B T.
\label{DL}
\end{equation}
Here, $k_B$ is the Boltzmann constant and $T$ the thermostat temperature. Equation (\ref{EOM-p2}) describes the atoms dynamic, which is affected by the spins motion and by the functionality of the exchange function $J(r_{ij})$.
 
 As shown in Eq.~(\ref{EOM_spin2}), the spin dynamics is modeled through the sLL equation. Here, $\bm{\omega}_i =- \frac{1}{\hbar} \frac{\partial \mathcal{H}_{\mathrm{mag}}}{\partial \bm{s}_i}$ is the effective field acting on spin $i$ and $\lambda_s$ is the spins damping parameter. $\bm{\zeta}(t)$ is the stochastic field which is also drawn from a Gaussian probability distribution with 
 \begin{eqnarray}
\langle\bm{\zeta}(t)\rangle &=& 0 
\; , 
\nonumber \\
\langle \zeta_a(t) \zeta_b(t') \rangle &=& 2 D_S \delta_{a b} \delta(t-t')
\; . 
\end{eqnarray}
In this case, the fluctuation-dissipation relation for the magnetic degrees of freedom is 
\begin{equation}
D_S=\frac{\lambda_s (1+\lambda_s^2) k_B T}{\hbar}. 
\label{DS}
\end{equation} 
In the Appendix, we formally derive Eqs.~(\ref{DL}) and (\ref{DS}). These are the parameter dependencies of $D_L$ and $D_S$ that allow for equilibration of the full system to a Boltzmann distribution $\propto e^{-\beta {\mathcal H}}$ with ${\mathcal H}$ in Eq.~(\ref{hamiltonian}), and the lattice and magnetic contributions specified below this equation.
 
We note that the Langevin equation presented in ref. \cite{tranchida2018massively}, Eq.~(\ref{EOM_spin2}),  does not have the stochastic field in the relaxation term (the last term) and, following ref.~\cite{tranchida2018massively}, we refer to it as the sLL equation (stochastic Landau-Lifshitz). When the stochastic field is added to both effective field terms the equation is usually called the sLLG equation (stochastic Landau-Lifshitz-Gilbert (sLLG). However, there is a small difference between the typical sLL equation and Eq.~(\ref{EOM_spin2}): the latter has the Gilbert factor included. Therefore, it is neither a typical sLL nor a typical sLLG equation. Still, once the noise parameters are well fixed, Eq.~(\ref{EOM_spin2}) also takes the system to thermal equilibrium at long enough times.
 
\subsection{\label{sec:simulation_details} Simulation details}

As it was described in the previous subsection, in the SLD  implementation of the software LAMMPS \cite{thompson2022lammps, tranchida2018massively, LAMMPS-spin}, the evolution of the system is described by two coupled Langevin equations, one for the spins and another one for the lattice degrees of freedom.

Each Langevin equation has a damping term and a random force (or field)  which are connected through the “fluctuation-dissipation" theorem, see Eqs.~(\ref{DL}) and (\ref{DS}). We have chosen the damping constants equal to $\lambda_s=0.5$ (for the spin degrees of freedom) and $\lambda_L= 1.0$ s$^{-1}$ (for the lattice). We use separate Langevin thermostats for the lattice and spin subsystems,
but both are set to the same temperature.
Most of the simulations were carried out at $T=10$ K but, as indicated in some cases, others were performed at higher temperatures.

The hysteresis loops were calculated applying an alternating magnetic field $\bm{H}$, and averaging the curves over up to ten different cycles. \textcolor{black}{At low temperatures, the individual loops exhibit a small dispersion of a few percent with respect to the average curve. A progressively larger dispersion is observed at higher temperatures, resulting in a dispersion of about 20\% for the magnetization in these cases. It is important to note that this variability, being inherent to the stochastic nature of the simulation method, does not alter the reported results nor the conclusions drawn from them. Instead, it reflects the complexity of the system and the method´s capacity to effectively capture temperature fluctuations, which provides valuable insights into the behavior of the magnetization dynamics across the temperature range studied.}

The field amplitude is varied discretely: it remains constant during certain simulation time, and then jumps to reach the value given by a function $H=H_{max} \cos(2 \pi f t)$. $H_{max}$ is set to be larger than the expected saturation field, and jumps do not have the same magnitude along the entire field range, given that the simulation time at each field value is kept constant. 

Typical MD simulations use a time step of 1 fs. However, to capture spin dynamics, the time step has been set in a range of 0.1 fs \cite{tranchida2018massively} to 10 fs \cite{westmoreland2020atomistic}. Here we use 0.1 fs but we have verified with several examples that using 1 fs does not change our results within the statistical spread. Each hysteresis cycle took around $2-8$ ns, giving MHz frequencies.
In particular, we have simulated field frequencies of $f=f_0$, $f_0/2$, $f_0/4$, and $f_0/8$, with $f_0 = 500$ MHz. These values are equivalent to sweeping rates of approximately $2.2 \times 10^9$ {\rm T}/{\rm s}, $1.1 \times 10^9$ {\rm T}/{\rm s}, $0.55 \times 10^9$ {\rm T}/{\rm s}, and $0.275 \times 10^9$ {\rm T}/{\rm s}, respectively, with ${\rm T}/{\rm s}$ representing Tesla per second. We note that in experiments, hysteresis loops are obtained using fields that change almost continuously, and the measurement of the magnetic moment of a sample can take up to several microseconds. Those time scales are well beyond the feasibility of SLD, or of other simulation methods like ASD, since typically simulations only reach nanosecond scale with sweep rates similar to ours \cite{bottcher2011atomistic, westmoreland2020atomistic,alzate2019optimal}. Nevertheless, as we show in the next section, the hysteresis loops quickly converge to a limiting curve as the frequency is decreased. 

Efficient minimization techniques, such as the one in Ref. \cite{ivanov2021minimize} would lead the system into efficiently finding and falling into the lowest energy minima, for every value of the external field, meaning that the spins would be aligned with the field as soon as the field direction is reverted and therefore would not produce a hysteresis loop. In addition, the use of efficient minimization techniques implies that there is no real time scale associated with the hysteresis loop frequency, which is important, for example, when calculating energy balance as in hyperthermia calculations.

Simulations were run for different angles, $\phi=0^{\circ}$, $45^{\circ}$, and $90^{\circ}$, between the external field $\bm{H}$ and the easy anisotropy axes. In practice, this was done by changing the uniaxial anisotropy axis directions while keeping $\bm{H}$ aligned in the $[001]$ direction ($z$ axis).  For uniaxial anisotropy, for instance, for $\phi=90^{\circ}$ or $45^{\circ}$  the unit vector $\bm{n}$ was oriented along the $[100]$ or $[101]$ direction. Most simulations where run with magnetic uniaxial anisotropy according to Eq.~(\ref{hamiltonian_uniaxial}).

For most cases we use cubic samples of dimensions $(10\times 10\times 10)a_0^3$ ($a_0 = 0.286$ nm is the lattice parameter of bcc Fe),  with periodic boundary conditions in all directions. The resulting system contains $2000$ atoms. To show that this size is large enough to capture the main hysteresis properties of the model at low temperatures, we also run a few simulations using a larger system with $(15\times 15\times 15)a_0^3$ cells (6750 atoms), as shown in the Supplemental Material (SM). We have also run a few simulations at higher temperatures, cases in which the magnetic fluctuations increase significantly. In these instances the volume of the system was set to $(32\times 32\times 32)a_0^3$ (around 65000 atoms) in order to avoid undesired finite-size effects.

Finally, we have to mention that the total magnetization $M$ as well as the components $M_x$, $M_y$ and $M_z$, will be expressed normalized to the ideal Fe bulk saturation magnetization $M_s$. The saturation magnetization is given by the maximum magnetic moment per unit volume. For bulk iron, in the volume of a bcc unit cell ($a_0^3$) there are 2 spins with magnetic moment $\mu = 2.2 \mu_B$ and, therefore, $M_s= 2 \times 2.2 \mu_B/a_0^3 \approx 1730$ kA/m. In this way, $M=1.0$ means $M=1730$ kA/m. 

\subsection{\label{sec:np_simulation_details} \textcolor{black}{Nanoparticles simulation details}}

\textcolor{black}{As it was mentioned in the introduction, we also performed hysteresis simulations of Fe NPs including lattice defects. In this subsection, we give additional details regarding these simulations.}

\textcolor{black}{We have run hysteresis simulations for two different nanoparticles, a pristine and a defective nanoparticle, at a temperature of $T=500$ K. The pristine NP was built simply by cutting a sphere with diameter 8 nm from the perfect bcc lattice. No further relaxation was carried out, to mimic usual frozen lattice simulations in atomistic spin dynamics.}

\textcolor{black}{The defective NP was built generating vacancies to the pristine NP above by removing $33.3 \%$ of the atoms, randomly. After this, an energy minimization was performed to relax the defective lattice.  The NP temperature was then raised to 900 K using a 20 ps linear ramp, held at 900 K during 20 ps, and then cooled-down to 500 K also with a linear ramp. After that, the temperature was held at 500 K during 20 ps,  resulting in a NP with a few small vacancy clusters, since most vacancies were absorbed at the surface. In order to include more extended defects, this NP was deformed mechanically using a flat rigid indenter \cite{amodeo2021indentation}, to mimic the strain that a NP could experience due to synthesis conditions. 
We used an indenter with a repulsive constant $K=20$ $\frac{eV}{\angstrom^3}$ ($\sim$ 3 TPa), typical of indentation in metals  \cite{ruestes2014nanoindent}. The indenter moved at 200 m/s, which is higher than typical velocities used in indentation simulations, but much lower than the sound velocity in the material and results are expected to be qualitatively similar. The NP is compressed up to a uniaxial strain of $\sim20\%$, but expands sideways in such a way that the compressive volumetric strain at the maximum indentation depth is lower than $5\%$. The indenter is then displaced upwards for unloading. The resulting NP was relaxed within the microcanonical NVE ensemble during 10 ps. The final configuration used for the calculation of hysteresis loops included small vacancy clusters and also a twin boundary, something expected in bcc NP \cite{hopper2020identification}, as seen in Figure~\ref{fig:NP_snapshot_pane} (f)-(g). The software Ovito \cite{stukowski2009visualization} was employed to render snapshots and to analyze the NPs microstructure and defects. Polyhedral Template
Matching (PTM) \cite{larsen2016robust} was used to obtain the crystal structures and surface mesh tool \cite{stukowski2014surface_mesh} was used to analyze the NPs surface and vacancies.}

\textcolor{black}{Vacancies imply a lower number of atoms that will produce a smaller saturation magnetization, $10\%$ smaller neglecting surface effects. NP topology changed from the roughly spherical shape of the pristine NP, including a more faceted surface, as shown in Figure~\ref{fig:NP_snapshot_pane} (h). There are ordered terraces and facets in the roughly spherical pristine NP. The defective NP shows surface disorder in Figure~\ref{fig:NP_snapshot_pane} (e), and large deviations from spherical shape in Figure \ref{fig:NP_snapshot_pane} (i). From separate tests, dislocations do not appear as stable for the NP size considered here. The magnetic moments are assumed equal to the bulk value for these NP simulations. We note that the magnetic moment is expected to vary near defects, and recent studies have explored variations near surfaces and vacancies~\cite{dosSantos2021, meyer2022influence}. However, these effects would be small and are not expected to change the overall behavior of the hysteresis loops.}

\textcolor{black}{For both NPs, the hysteresis simulations were carried out at $T=500$ K, considering uniaxial magnetic anisotropy (Eq.~(\ref{hamiltonian_uniaxial})) with an anisotropy constant $K_1=35$ $\mu$eV/atom which has been considered previously for Fe NPs \cite{ibusuki2001magnetic, pastor1995magnetic}. The external magnetic field is applied at zero degrees with respect to the anisotropy axis and it is varied at a sweep rate of $SR=0.227 \times 10^8$ T/s.} 

\textcolor{black}{In these NP simulations the exchange function $J(r_{ij})$ (Eq.~(\ref{exchange_dep})) is fitted to the ab-initio data by Pajda et al.~\cite{pajda2001ab} using the following fitting parameters:
$\alpha = 25.498$ meV, $\gamma = 0.281$, $\delta = 0.1999$ nm. Previous simulations of Fe NPs with these parameters have shown excellent agreement with experimental observations  \cite{dosSantos2020} and we note that for this parametrization of $J(r_{ij})$, the Curie temperature for an 8 nm NP is close to $600$ K. The remaining parameters are the same as those of the bulk simulations. 
}


\section{\label{sec:results} Results and discussion}

In this section, we first analyze how the computational parameters, 
such as the simulation run time and the field frequency, affect the final outcome of the calculations. The aim is to establish optimal parameters that minimize the computational cost without affecting the reliability of the calculations. Next, we focus on studying how the hysteresis loops can change when different physical properties like anisotropy, exchange interaction, and damping parameter, are modified. \textcolor{black}{These studies are carried out on the bulk samples described in the methods section}. In a first stage, we analyze these effects without coupling the lattice vibrations to the spins dynamics, i.e. we consider for most cases spin dynamics simulations with the atoms fixed at their ideal-lattice positions. The influence of coupling to the lattice vibrations is analyzed in the final part of this section, where we run full SLD simulations \textcolor{black}{for both bulk and nanoparticles samples}.

\subsection{Simulation time}
We start analyzing the effect of the simulation run time $t_{\rm sim}$. 
This quantity represents the interval during which the external field $\bm{H}$ 
is held constant before increasing or decreasing its absolute value by a given amount. It is basically the total time of simulation used to calculate each individual discrete point of a hysteresis loop. After changing the external field, the magnetic moments must relax to a new stable configuration and therefore $t_{\rm sim}$ must be large enough to allow this process to occur. It is important, then, to determine an optimal value of this quantity that allows to achieve this goal. 

\begin{figure}[t]
\centering
\includegraphics[width=0.99\columnwidth]{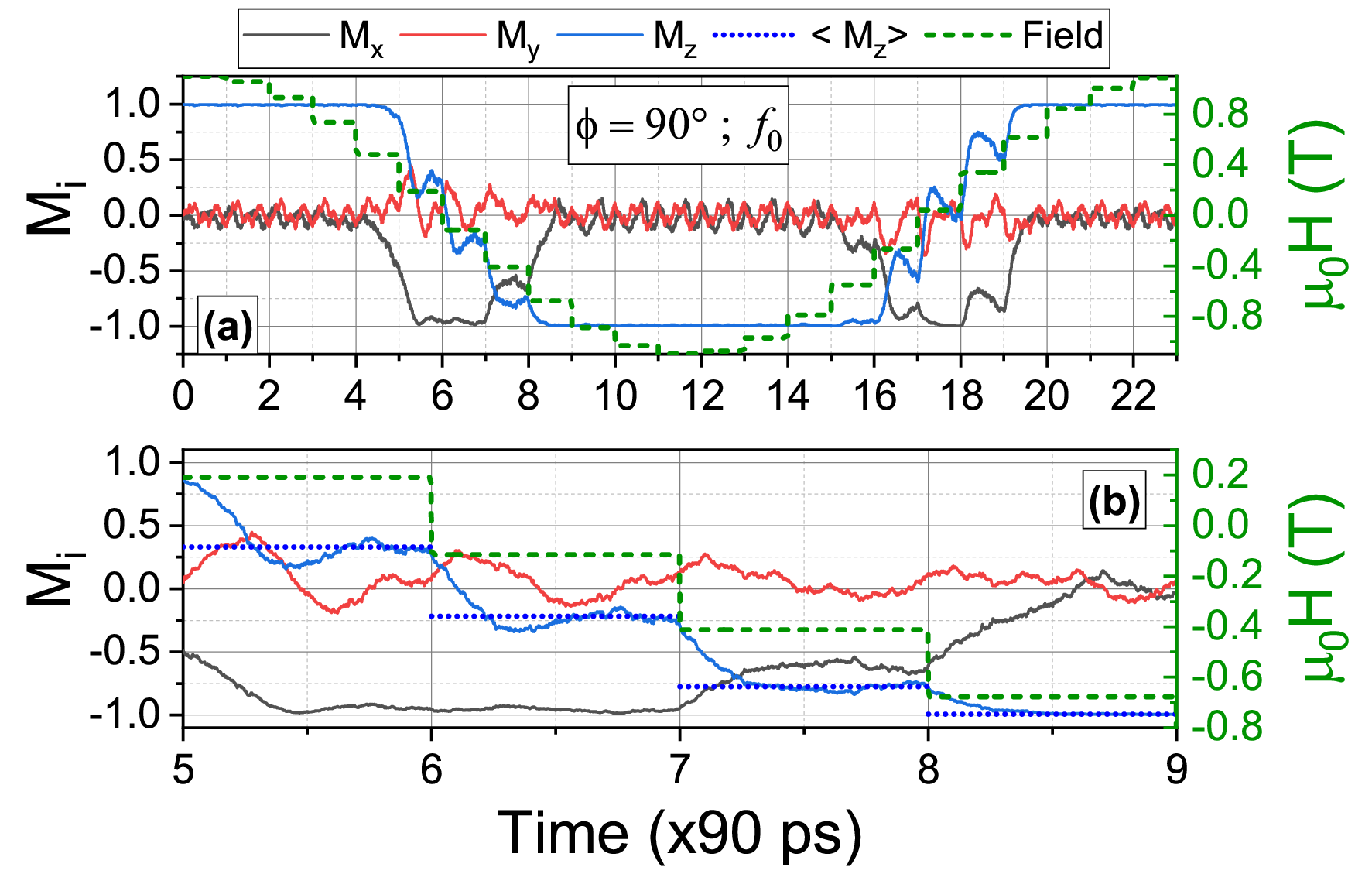}

  \caption{(Color online.) Time evolution of the components of the normalized total magnetization 
throughout an entire cycle (a) and half a cycle (b). 
The field value at each time is also included (green dashed line and right axis). 
Note that each field value is kept constant for 90 ps ($9\times 10^5$ steps) of simulation time. 
In (b), the time average value of $M_z$, $\langle M_z \rangle$ (blue dotted line), obtained from the last 30 ps of simulation, is also added for comparison.} 
\label{fig:M_vs_time_90}
\end{figure}

In principle, one can imagine an individual spin precessing around the field direction, with a frequency which increases with the field magnitude. For Fe, the resulting Larmor frequency gives a period of 36~ps for a 1.0~T field, and several precession periods are needed to describe the magnetization evolution. However, for damped dynamics, the frequency remains the same but the spin spirals down towards the field direction, allowing shorter simulations \cite{evans2014atomistic}. Figure~\ref{fig:M_vs_time_90} shows the time evolution of the components of the normalized total magnetization along the three Cartesian axes, $M_x$, $M_y$, $M_z$, corresponding to directions $[100]$, $[010]$, and $[001]$, respectively, during the simulation of an entire loop for the case $\phi=90^{\circ}$. Unless otherwise stated, for this and subsequent calculations 
we use uniaxial anisotropy. As it can be noticed by inspection of this figure, 
the applied field (green dashed curve and right axis) is maintained at a constant value for 90 ps before increasing or decreasing it to the next value, so $t_{\rm sim}=90$ ps. We are interested here in the stabilization of the magnetization along the field direction, $M_z$. 
During the first moments after the field value is varied a fluctuation period is observed, specially in the region around the switching field, before $M_z$ reaches a stable value  after approximately 45 ps. This is better appreciated in Figure~\ref{fig:M_vs_time_90}(b), where a zoom in the region around the first magnetization switching is displayed. In this figure, we have also included the average value of $M_z$, $\langle M_z \rangle$, which is obtained in our simulations from the last 30 ps of each step, where a well stabilized magnetization is observed. We note that $t_{\rm sim}=90$ ps is similar to simulation times used in micromagnetic and ASD calculations of hysteresis loops \cite{aurelio2020hysteresisCore-shell,behbahani2020coarse,behbahani2021multiscale,bottcher2011atomistic,westmoreland2020atomistic,alzate2019optimal}. We have checked that longer simulation times do not significantly affect the final results, as it is shown in Figure~\ref{fig:Cycle_vs_time} in the SM. 
It is crucial that our magnetization dynamics is well described near the switching field values. For fields of 0.5 T, the precession period is $\sim 72$ ps, and one would need a few ns of simulation time for low damping. However, thanks to the high damping values discussed below, shorter simulation times can be employed. 

\begin{figure*}[ht]
\centering
\includegraphics[width=0.99\textwidth]{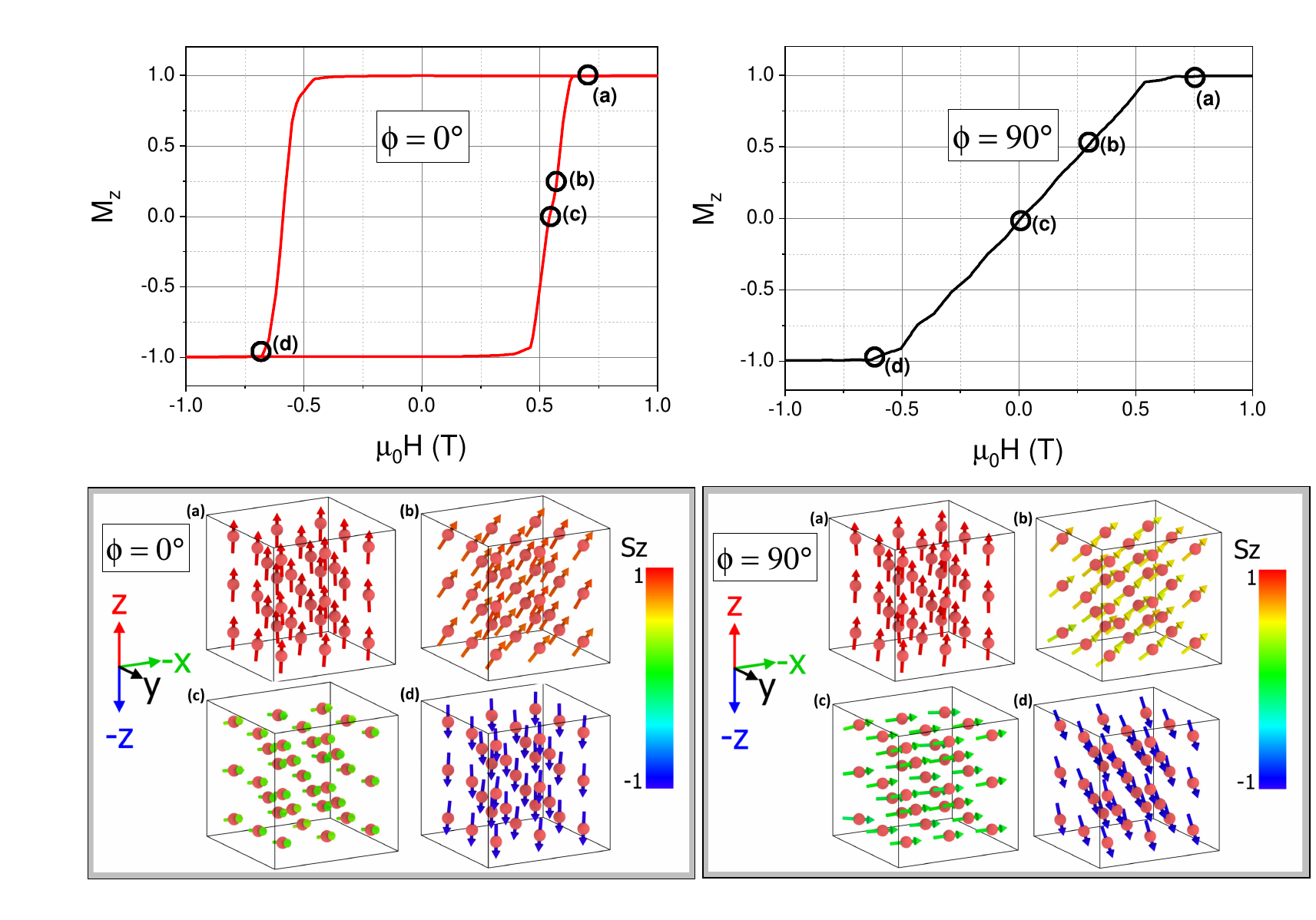}
  \caption{(Color online.) Top images: resulting hysteresis loops obtained at the converged frequency for $\phi=0^{\circ}$ (top left) and $\phi=90^{\circ}$ (top right). Bottom images: snapshots of a fraction of the system, showing typical spin configurations at different stages of the hysteresis loop for the case $\phi = 0^{\circ}$ (bottom left) and $\phi = 90^{\circ}$ (bottom right). Each snapshot, (a), (b), (c) and (d), corresponds to the points marked on the hysteresis curves (upper images). In the snapshots, the atoms are represented as red spheres and the spins are colored according to their orientation along the field direction $z$. Each cubic cell represents a fraction of the system with a volume of (8 \AA)$^3$, extracted from the center of the original system.} \label{fig:snapshot_panel}
\end{figure*}

\begin{figure}[ht]
\centering
\includegraphics[width=0.99\columnwidth]{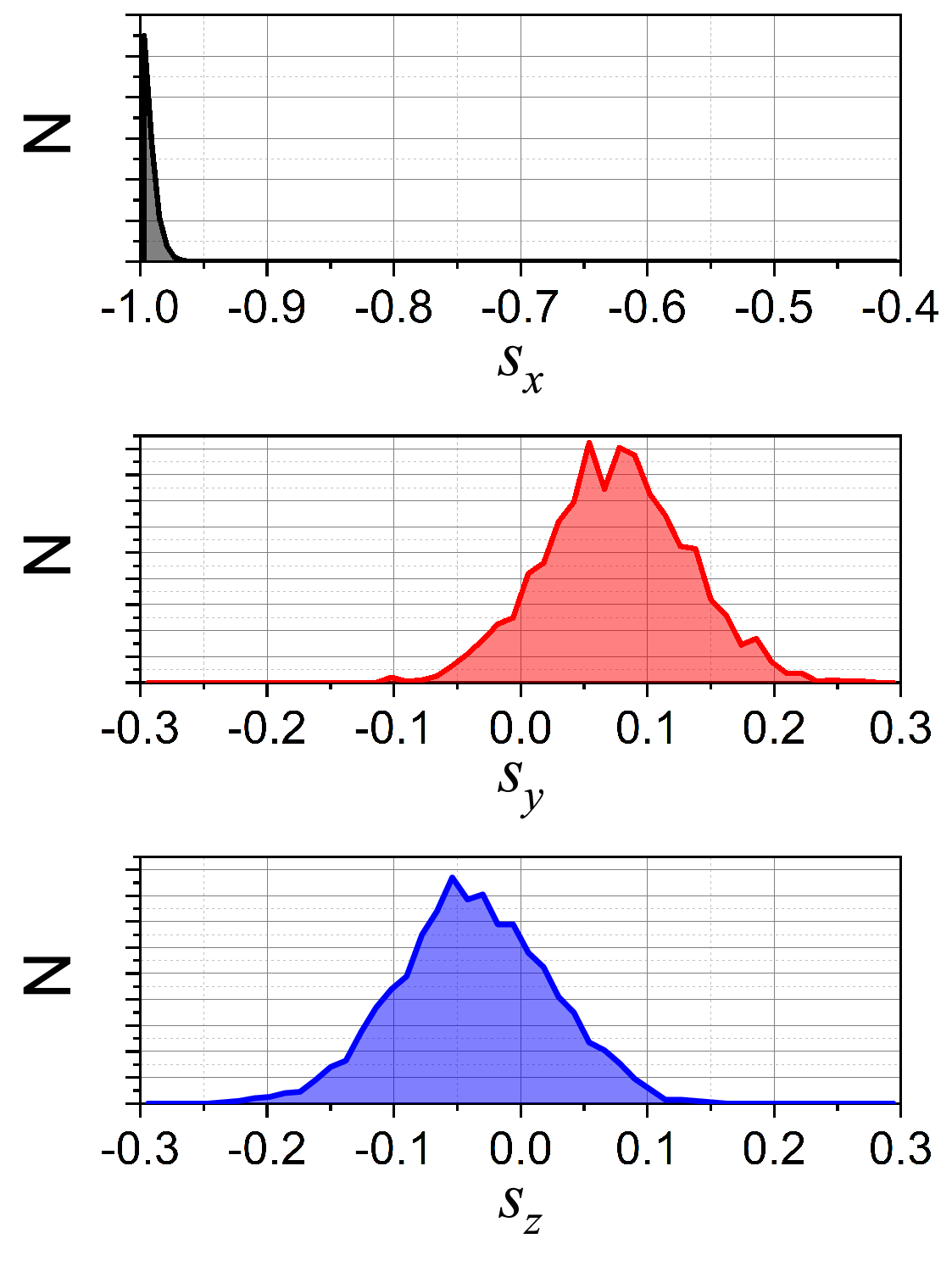}
  \caption{(Color online.) Histograms of individual spins orientations $s_x$, $s_y$, $s_z$ along the three axes. The histograms correspond to the state of the system described in the snapshot~(c) of Figure~\ref{fig:snapshot_panel} for the case $\phi = 90^{\circ}$. 
 }
\label{fig:histograms}
\end{figure}

Using $t_{\rm sim}=90$ ps and frequency $f_0/4$, we calculate the hysteresis loops for the cases $\phi=0^{\circ}$ and $\phi=90^{\circ}$. These curves, along with snapshots showing typical spin configurations at different stages of the process, are shown in Figure~\ref{fig:snapshot_panel}.
As we can see, the system behaves qualitatively according to what the SW model predicts, i.e. all spins are roughly in sync like a single macrospin \cite{stoner1948mechanism,tannous2008stoner}.
This is expected for low-temperature simulations, but deviations would occur at higher temperatures. 
Still, Figure~\ref{fig:histograms} shows that there is a roughly Gaussian distribution of spin values, 
an appreciable deviation from the simpler macrospin assumption. The spin orientation histograms in this figure correspond to the state of the system described by the snapshot (c) of the $\phi=90^{\circ}$ case of Figure~\ref{fig:snapshot_panel} (bottom right). 

Another observation is the behavior of the magnetization components around $H \approx 0$
(being $H$ the modulus of $\bm{H}$) for the case 
$\phi=90^{\circ}$. In Figure~\ref{fig:M_vs_time_90} we can see that when this field takes values close to zero, $M_z \approx 0$ (and also $M_y \approx 0$) while $M_x$ takes values close to saturation. This means that, at this stage, all spins are aligned approximately along the $x$ direction (the easy axis of magnetization for the uniaxial anisotropy), showing that the simulation reproduces well the expected behavior for $\phi=90^{\circ}$.
The snapshots in Figure~\ref{fig:snapshot_panel} (right panels) confirm this interpretation. 

Since these tests confirm that our simulations are consistent with some general physical properties of the system, from now on we set $t_{\rm sim}=90$ ps. 

\subsection{Hysteresis loops and the effect of field frequency}

\begin{figure}[ht]
\centering
\includegraphics[width=0.99\columnwidth]{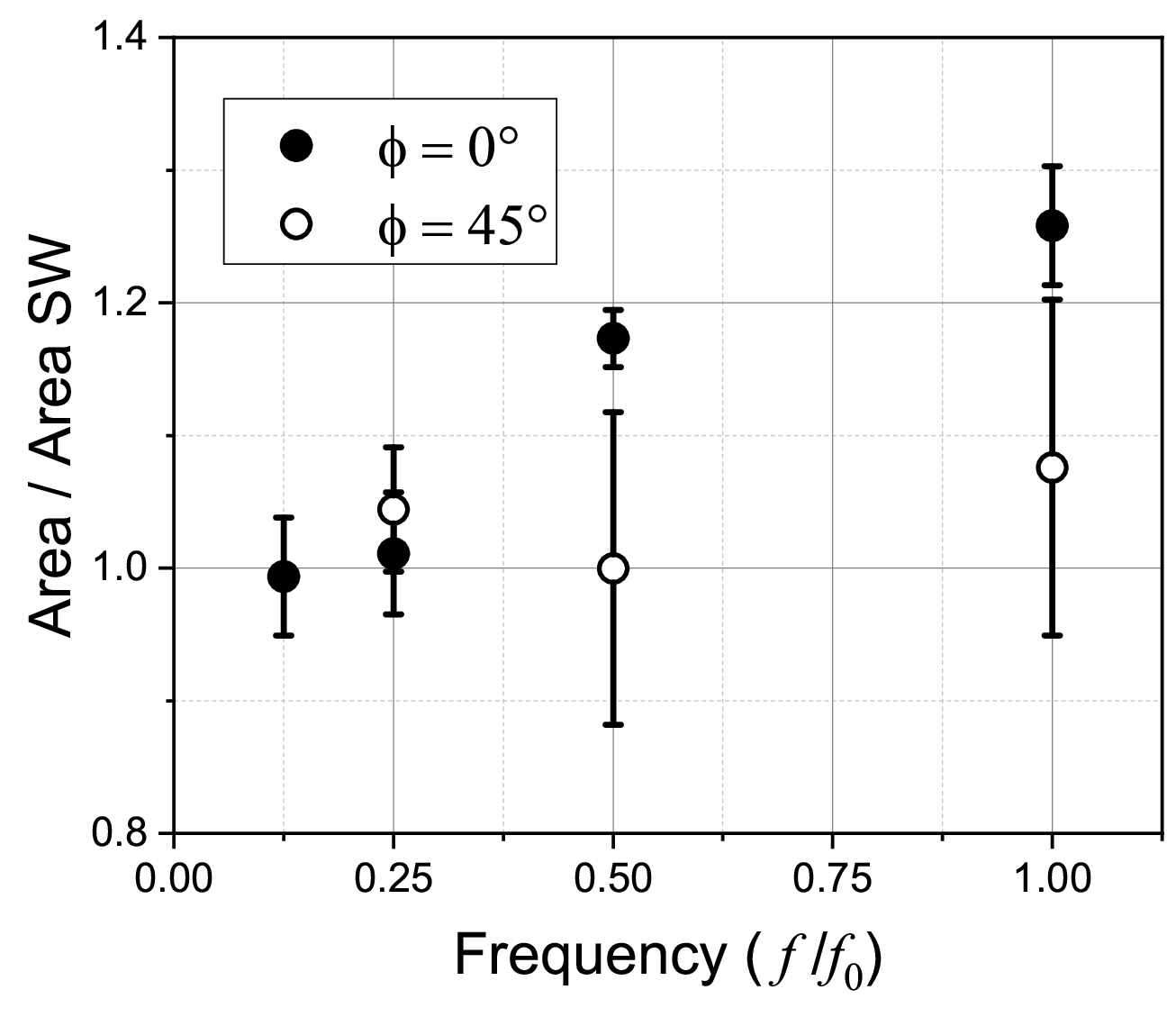}
  \caption{(Color online.) Loop area vs field frequency for the cases with 
$\phi=0^{\circ}$ and $\phi=45^{\circ}$. 
The values of the area are normalized by the area of the SW loop.} 
\label{fig:area_vs_freq}
\end{figure}

As it was argued in Section \ref{sec:methods},  it is not possible to calculate hysteresis loops at typical very low experimental frequencies using SLD simulations, due to intrinsic time limitations of atomistic simulation approaches. Instead, we are limited to working with high frequencies in the order of MHz. 
However, high frequency hysteresis loops appear to converge to a limit curve as the frequency is decreased and, as the following analysis shows, this limit loop should not be so different 
from those measured experimentally or calculated theoretically. Following this line of argumentation, we calculate the hysteresis loops for different frequencies. In Figure~\ref{fig:area_vs_freq}, we show how the area of the hysteresis loops depends on the field frequency for the cases with $\phi=0^{\circ}$ and $\phi=45^{\circ}$. Note that we exclude the case $\phi=90^{\circ}$ from this analysis, since for this angle the curves do not present an hysteretic behavior. The curves in Figure~\ref{fig:area_vs_freq} exhibit a typical crossover from a high frequency dynamic regime (for which the area of the loops is large) to a low frequency dynamic regime (a weak variation of the area with field frequency) \cite{Dolz2020}.

Given that the simulation time for a given field value is always the same and, in principle, long enough to cover spin relaxation for high damping, higher frequency would only mean larger changes in the discrete field values. The larger the field jumps, the larger the change across the energy landscape, and the system might not adapt fast enough within the 90 ps of simulated time, decreasing spin switching probability and generating wider loops. The magnitude of the jumps around the region of the switching field are $\sim 0.3$ T for $f_0$ and $\sim 0.06$ T for $f_0/4$. Therefore, the stabilization of the frequency for $\phi=0^{\circ}$ occurs when the field jumps are about one order of magnitude lower than the value of the converged coercive field.

For $\phi=45^{\circ}$, we found convergence at a frequency of $f_0/2$ while for the $\phi=0^{\circ}$ case it is not until a frequency of $f_0/4$ that the loop area reaches a stable value. At these frequencies the area of the loops coincides (within the statistical errors) with the predictions of the SW model. Figure \ref{fig:loop_convergence} in the SM shows the corresponding hysteresis curves calculated at several different frequencies for the case $\phi=0^{\circ}$, evidencing convergence as frequency decreases. This is similar to results showing convergence as loop time increases in ASD simulations \cite{chureemart2017hybrid} for simulation times similar to our SLD runs. This analysis shows that, although the convergence frequencies found in this study are much higher than those typically used in experiments, they are low enough to let the magnetization equilibrate with the field direction at each field step. This means that, at these rates, our simulated hysteresis loops should not present large discrepancies with experiments, as discussed for example by Westmoreland {\it et al.} \cite{westmoreland2020atomistic}.

\begin{figure}[ht]
\centering
\includegraphics[width=0.99\columnwidth]{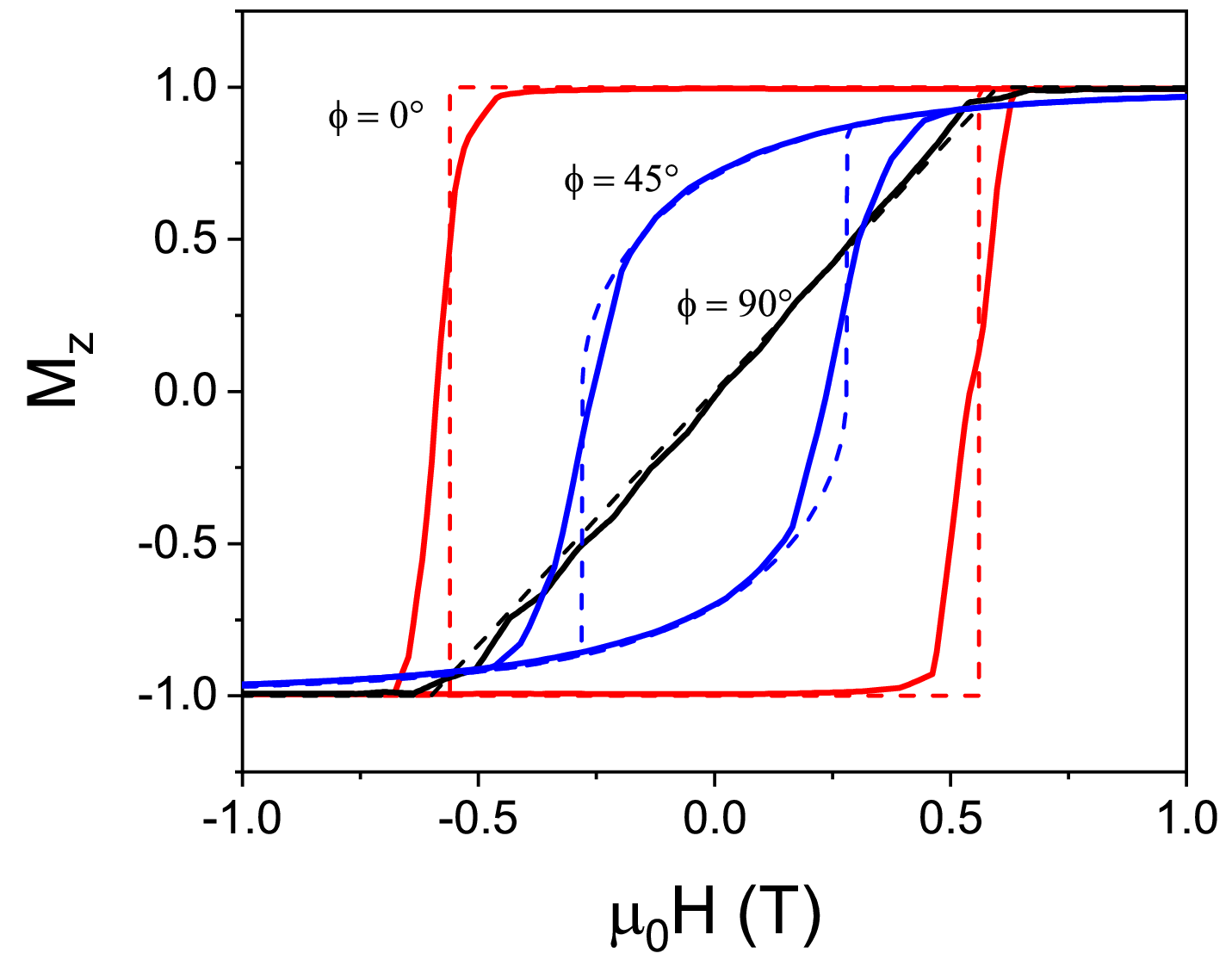}
  \caption{(Color online.) Hysteresis curves calculated with SLD simulations (full lines) at different angles $\phi$ compared to the theoretical ones according to the SW model (dashed lines). The SLD curves correspond to the average curve obtained from many different cycles. In all cases a field frequency $f=f_0 / 4$ was used.} \label{fig:SLD_vs_SW}
\end{figure}

In Figure~\ref{fig:SLD_vs_SW} we compare the hysteresis loops calculated for 
$\phi=0^{\circ}$, $\phi=45^{\circ}$, and $\phi=90^{\circ}$ at a single frequency $f_0/4$, with those of the SW model \cite{cullity2011introduction}. As we can see, a good agreement is observed in all cases. Nevertheless, it is important to stress that we do not expect that the simulation curves should fit exactly the theoretical ones. Unlike the SW model (for which the macrospin approximation is considered at $T=0$), the SLD simulations are carried out at low but finite temperature for many magnetic atomic moments which, along a hysteresis cycle, do not rotate coherently in a perfect way (see Fig.~\ref{fig:histograms}). For example, for the $\phi=0^{\circ}$ case, we do not obtain a rectangular curve because thermal fluctuations and also the existence of many degrees of freedom,
make it easier for the magnetization to change its orientation at an external field value slightly lower than predicted by the SW model.

\begin{figure*}[ht]
\centering
\includegraphics[width=1 \textwidth]{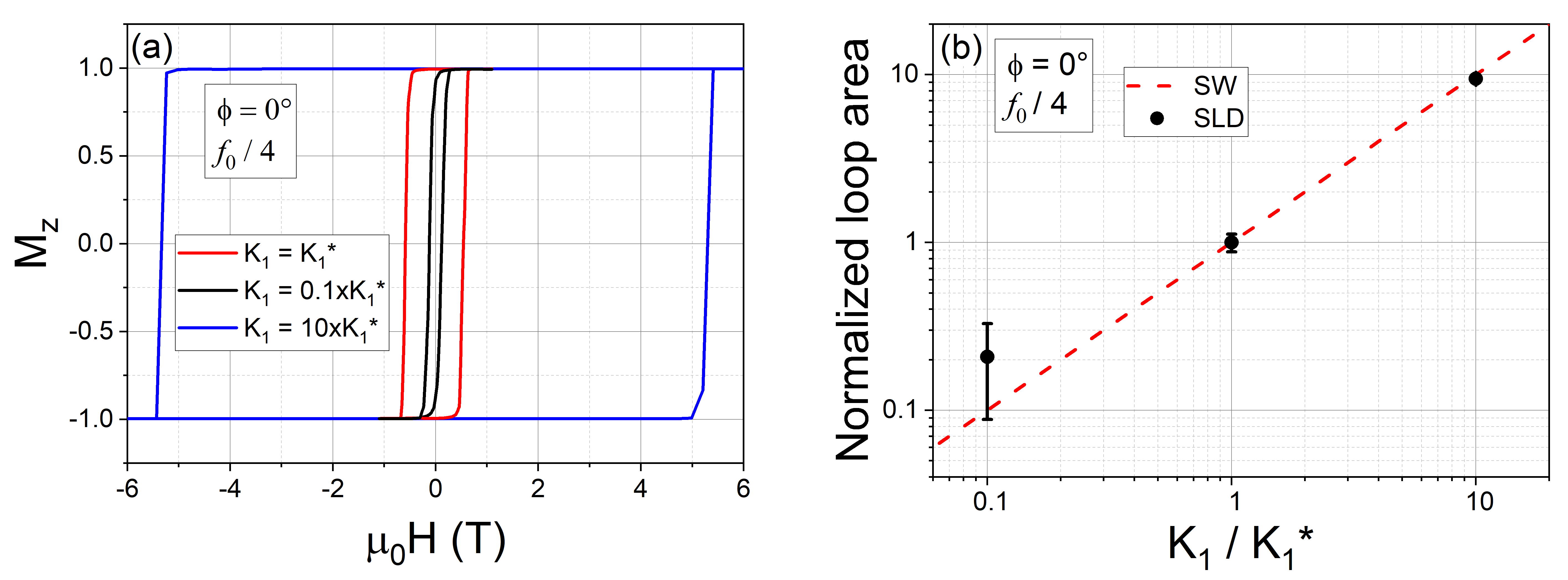}
\caption{(Color online.) (a) Loops for different values of uniaxial anisotropy $K_1$. 
(b) Loop area normalized to the area of the loop with $K_1^*=35$ $\mu$eV/atom, 
against $K_1/K_1^*$.  The dashed line is the prediction of the SW model. } 
\label{fig:area_vs_aniso_0}
\end{figure*}


\subsection{Anisotropy}

According to the SW model, the coercive field and, therefore, 
the area of the hysteresis loop when the external field is applied along the easy axis ($\phi=0^{\circ}$), is proportional to the anisotropy constant $K_1$ in Eq.~(\ref{hamiltonian_uniaxial}). We have explored this dependency in our SLD simulations by reducing and increasing the original anisotropy constant value (throughout this test called $K_1^*=35$ $\mu$eV/atom) by a factor of 10. Figure~\ref{fig:area_vs_aniso_0} (a) shows that the resulting simulated loops qualitatively exhibit the expected behavior, that is, the loop broadens (narrows) according to the increase (reduction) of the anisotropy constant. This is further confirmed in Figure~\ref{fig:area_vs_aniso_0} (b)  where it can be seen that the loop area follows an approximately linear behavior as a function of $K_1/K_1^*$, deviating slightly from the SW model prediction only for small values of this constant. These deviations arise from the combination of two effects. On the one hand, this is due to the fact that simulated loops are ``rounded'' near the point of 
magnetization reversal, deviating from perfect rectangular loops as explained above. The discrepancy with the SW model area becomes increasingly important as the loop narrows. On the other hand, as it was mentioned in the methods section, large anisotropy helps stabilizing the magnetization of the system \cite{alzate2019optimal}. For the case of low anisotropy ($K_1=0.1K_1^*$) there is poor stabilization of $M_z$ during the 90 ps of simulation of each field step (specially near the region of magnetization switching), as seen in 
Figure~\ref{fig:Equil_vs_aniso}.  Much longer simulation times would be required for low anisotropy values. Given that our smallest anisotropy is the Fe bulk anisotropy, we could compare with some experimental results. Magneto-optic Kerr effect measurements of Fe(100) appear to give a slightly lower coercive field than our simulation estimate \cite{bansmann1992surface},  although those measurements were obtained at higher temperatures and for multi-domain samples.

\begin{figure}[htb]
\centering
\includegraphics[width=0.99\columnwidth]{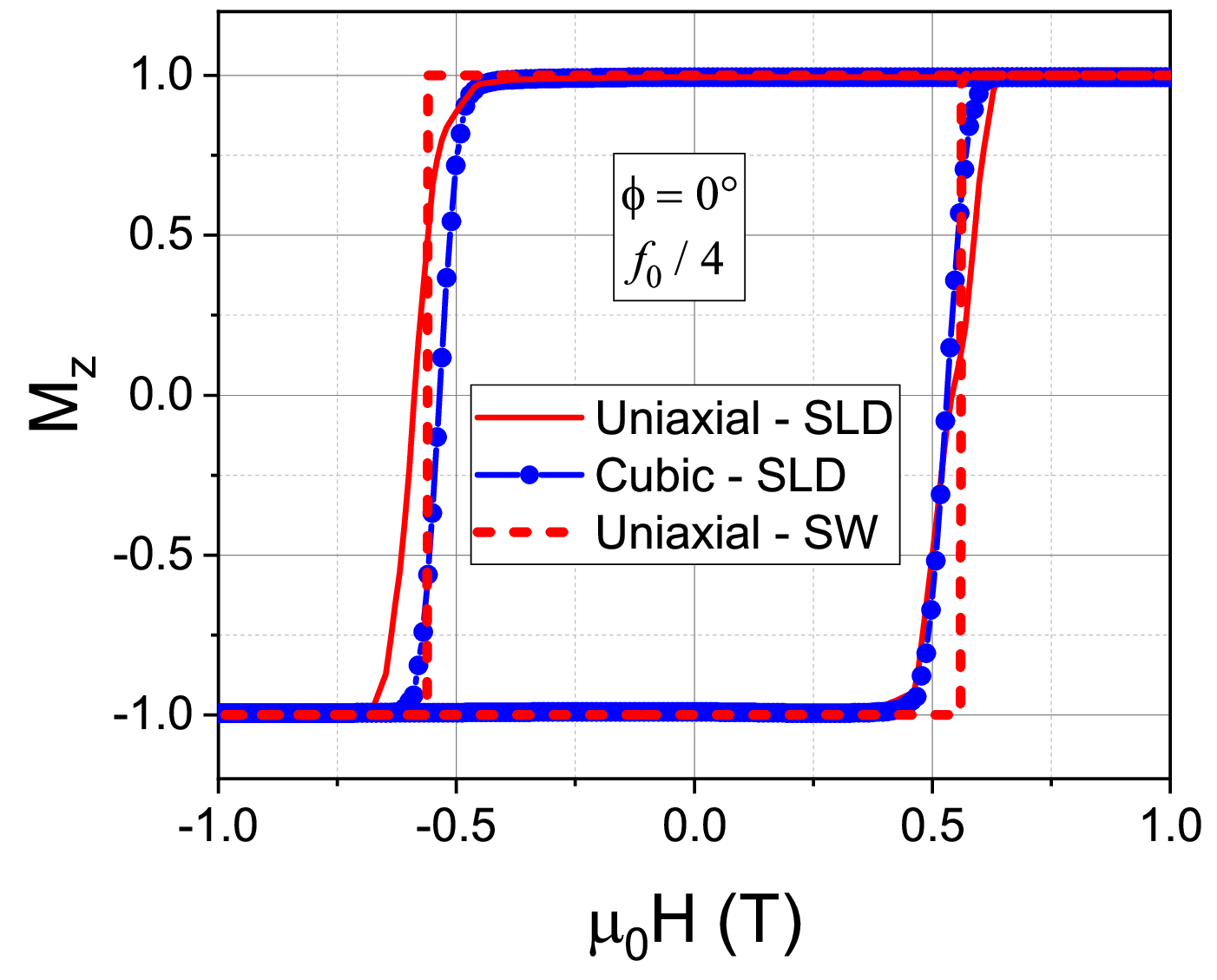}
  \caption{(Color online.) Hysteresis loops for uniaxial (red full line) and cubic (blue circles) 
anisotropies for the $\phi=0^{\circ}$ case and frequency $\nicefrac{f_0}{4}$. The results are compared to the SW prediction for uniaxial anisotropy (red dashed line).} \label{fig:cubic_vs_uniaxial}
\end{figure}

We have also analyzed the influence of anisotropy symmetry.
Loops calculated with SLD simulations for systems with uniaxial and cubic anisotropies, Eqs.~(\ref{hamiltonian_uniaxial}) and~(\ref{hamiltonian_aniso}), are shown in Figure~\ref{fig:cubic_vs_uniaxial} for the case $\phi=0^{\circ}$. We compare these curves to the corresponding one for the SW model with uniaxial anisotropy. The reason is that at $T=0$ K both simulation curves should coincide with the latter. Our results show that for $\phi=0^{\circ}$ the hysteresis loop does not depend on the type of anisotropy. Therefore, the small differences between the curves shown in Figure \ref{fig:cubic_vs_uniaxial} can be attributed to the effect of temperature, which is more pronounced in the case of cubic anisotropy. The area of the loop is slightly smaller for cubic anisotropy, similar to the results reported in Ref.~\cite{aurelio2020hysteresisCore-shell}. Usov and Peschany \cite{usov1997theoretical} considered the SW model for uniaxial and cubic anisotropies, for randomly oriented nanoparticles. They report that the maximum normalized coercive field in both cases is 1, in agreement with our result. For a random collection of anisotropy orientations, the resulting coercive field for cubic anisotropy was $\sim 0.68$ of the one for uniaxial.

\begin{figure}[ht]
\centering
\includegraphics[width=0.99\columnwidth]{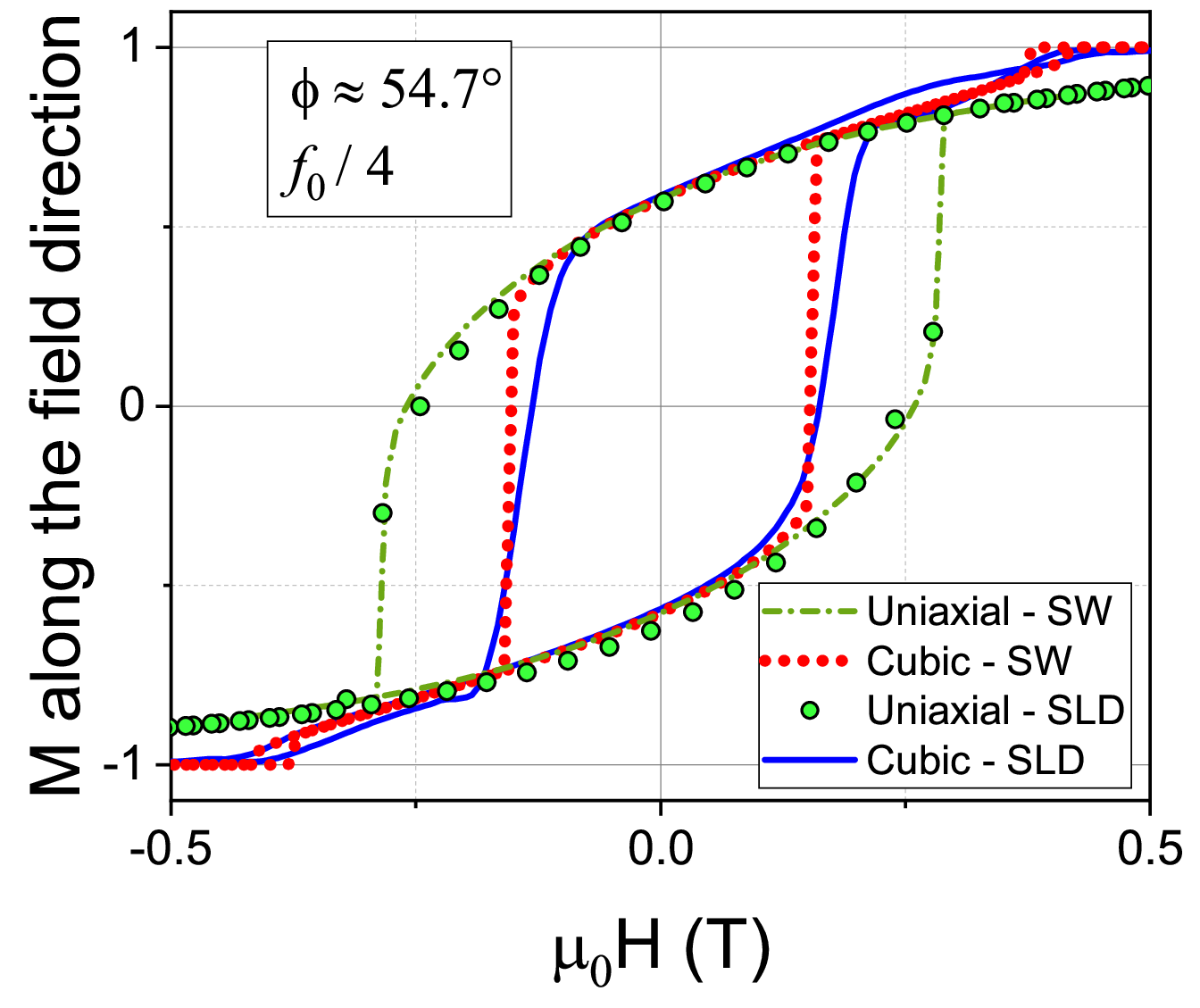}
  \caption{(Color online.) Hysteresis loops for an external field applied along the [111] direction ($\phi \approx 54.7^{\circ}$) and frequency $\nicefrac{f_0}{4}$, for cubic and uniaxial anisotropy. The loops obtained with SLD are compared to the predictions of an extension of the SW model that incorporates cubic anisotropy (red short-dotted line), and to the SW with uniaxial anisotropy (green dot-dashed line).} \label{fig:cubic_vs_uniaxial_45}
\end{figure}

It is more interesting to analyze the difference between uniaxial and cubic anisotropies, when the field is applied in the $[111]$ direction. For this field orientation, we consider a SW model, the Hamiltonian of which has an arbitrary anisotropy term. We start by applying a strong enough external field such that 
the Hamiltonian has a single minimum with the magnetization $\bm{M}$ pointing in the direction of $\bm{H}$. Then, we decrease the intensity of the field using small jumps. At each step we use a steepest descent method to try to escape from the energy minimum, so that if it becomes unstable, a new minimum can be reached. Using this simple algorithm, we have simulated the zero temperature dynamics of the SW model with both uniaxial and cubic anisotropies. As shown in Figure \ref{fig:cubic_vs_uniaxial_45}, in this case the corresponding hysteresis loops for the SW model are very different. However, as expected, the loops calculated with the SLD simulations agree very well with the curves for the SW model for both cubic and uniaxial anisotropies.  In the case of cubic anisotropy, while the remanent magnetization for both SW and SLD curves is $\pm 1/\sqrt{3}$, there is a qualitative difference between them for field values greater than the coercive one: a small hysteresis behavior present in the loop of the SW model is not well reproduced in the SLD simulation. The reason for this is that the energy barrier separating these states is of approximately $0.3$ $\mu$eV/atom, several orders of magnitude smaller than $k_B T \sim 8.6 \times 10^{-3}$ eV. Only a simulation at much lower temperature would be able to reproduce this behavior. The cubic case has a lower coercive field, which could be somewhat expected from results for a collection of nanoparticles with random orientation \cite{usov1997theoretical}.

To conclude this section, we mention that for a couple of cases, we have tested a lattice-dependent Neel´s anisotropy within the framework of Nieves {\it et al.} \cite{nieves2021neel} and the results obtained are similar to the ones for uniaxial or cubic anisotropy presented above.

\subsection{Exchange interactions}

\begin{figure}[ht]
\centering
\includegraphics[width=0.99\columnwidth]{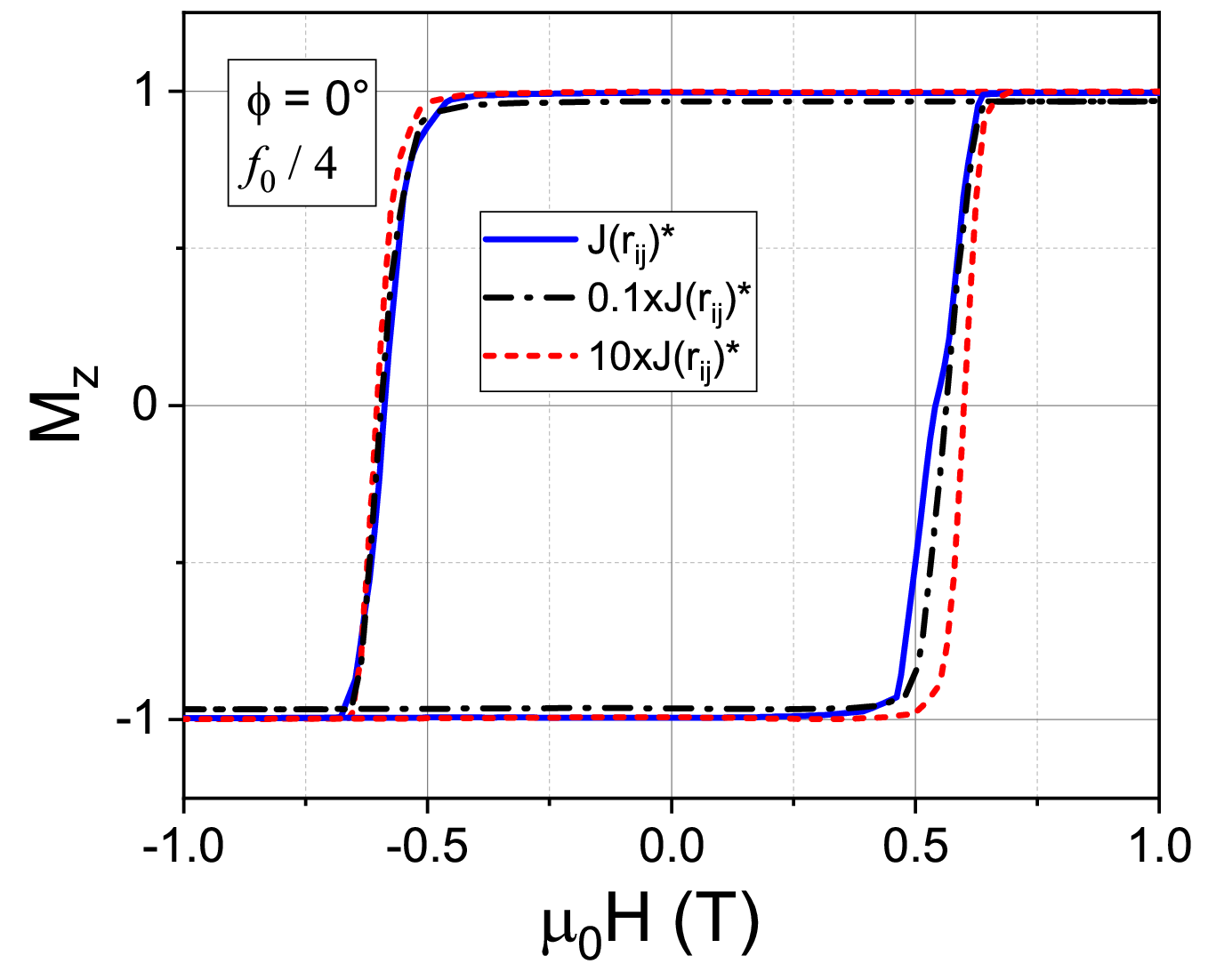}
  \caption{(Color online.) Simulated hysteresis curves for the case $\phi=0^{\circ}$ considering different magnitudes of the exchange coupling.} \label{fig:Jx10}
\end{figure}

The exchange energy does not play a role in the SW results, and within that framework there should be no effect of the exchange interaction on the simulated hysteresis loops. To test this prediction within the SLD framework, we have run simulations with different values of the function $J(r_{ij})$. In Figure~\ref{fig:Jx10}, we compare the curves obtained using the original exchange $J(r_{ij})^*$, with the ones obtained by increasing or reducing this interaction by a factor of 10. As it can be seen, loops are nearly unaffected by these changes. Increasing the magnitude of the exchange interaction is expected to give higher  magnetization 
values and broader loops \cite{kachkachi2000surface}, but 
at low temperatures (as in this case) this effect is expected to be very small.  Larger differences should be noticed at higher temperatures where thermal fluctuations start to play a major role 
in the magnetization behavior. Nevertheless, even at these low temperatures, 
a few subtle effects can be noticed: 
(i) for the simulation with $J(r_{ij})=0.1\times J(r_{ij})^*$ a small but consistent reduction of the saturation magnetization is observed, and 
(ii) for $J(r_{ij})=10\times J(r_{ij})^*$ the loop becomes more rectangular. This last effect is also observed in other theoretical approaches based on the SW model
which include exchange interactions through a mean field approach as in Refs.~\cite{atherton1990_SW_meanField, zhang2003magnetization}.
Lower/higher exchange will lead to lower/higher Curie temperature ($T_C$) and spin ordering will be modified. For the same value of the external field, the histogram of spin values is narrower and with a higher mean value for the higher exchange, facilitating the macrospin behavior, leading to a larger saturation field and helping with the sudden spin flip which causes a more square loop.
\subsection{Damping parameter}
The value of the Gilbert damping parameter in the sLL equation is usually chosen to be in the range $0.01-1$. Experimental estimates for Fe are around 0.001. However, the possibility of increasing the spin damping values for numerical convenience is often explored. In SLD simulations a value of $\lambda_s=0.1$ was used in Ref.~\cite{dosSantos2020}. ASD \cite{evans2020atomistic} and micromagnetic \cite{aurelio2020hysteresisCore-shell,behbahani2020coarse} simulations use values in the range $0.1-1$. 

\begin{figure*}[htb]
\centering
\includegraphics[width=0.99\textwidth]{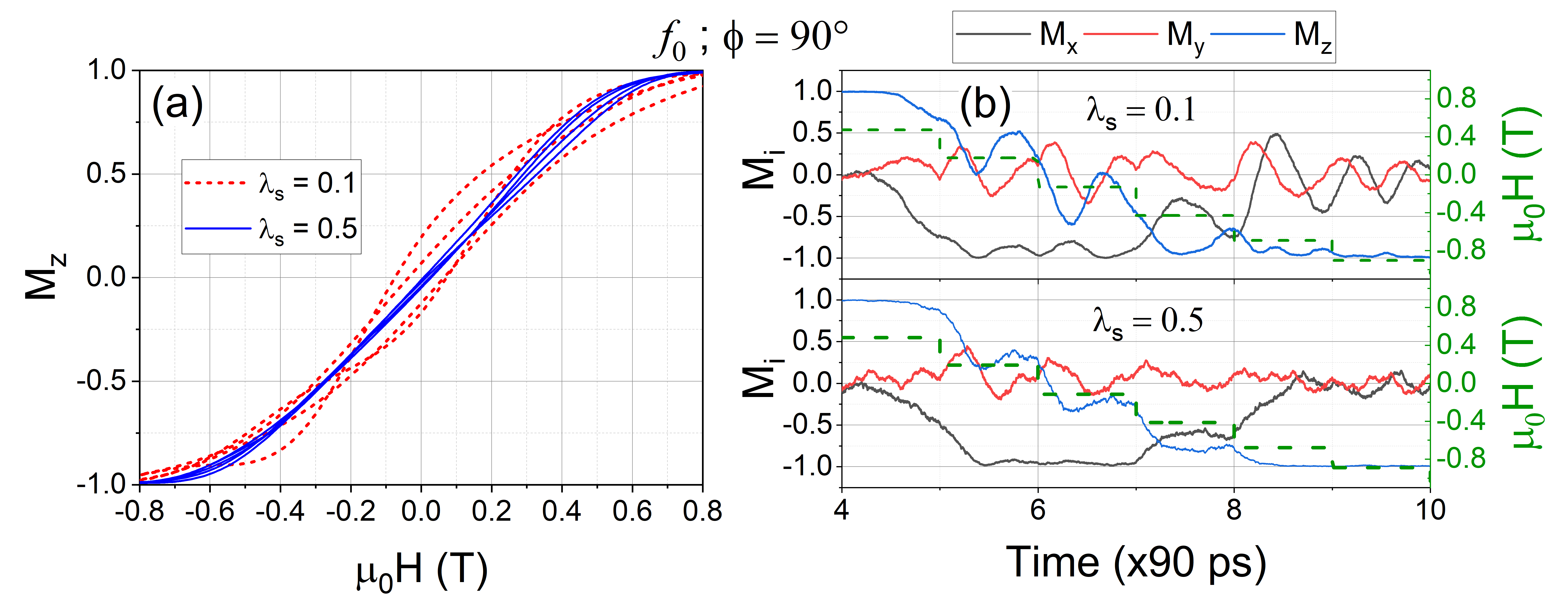}
  \caption{(Color online.) (a) Hysteresis curves for the case $\phi = 90^{\circ}$ with a field frequency $f_0$ and damping parameters $\lambda_s=0.1$ (dashed line) and $\lambda_s=0.5$ (full line). The plot shows several cycles for each case. (b) Time evolution of the components of the magnetization throughout the simulation of half a cycle for the cases considered in (a). The field value at each time is also included (green dashed line and right axis). Note that each field value is kept constant for 90 ps ($9\times 10^5$ steps) of simulation time.} \label{fig:damping}
\end{figure*}

The dynamic of a single spin in a magnetic field is well established, and there are studies discussing the role of damping and simulation times \cite{evans2014atomistic}. However, obtaining spin relaxation times at a given damping for correlated multi-spin systems at finite temperature will typically require simulations. In general, as the damping increases, there is a faster energy loss, and the relaxation time required for the system to reach a steady state decreases, with spins aligned with the preferred orientation determined by the external field and the anisotropy. Therefore, it is possible to use higher frequencies to calculate the hysteresis loops, which become narrower as damping grows. In the same spirit, in most of the simulations presented here we use a larger damping value, $\lambda_s =0.5$. Using lower damping is possible but would require lower frequencies as discussed below. The parameter $\lambda_s$ is related to the optimal value of the simulation run time $t_{\rm sim}$. 

In micromagnetic simulations, it has been proposed to use both very high damping and high loop sweep rates (SR),
in order to achieve computational efficiency \cite{behbahani2020coarse}. The rationale behind this proposal was as follows. The magnitude of the coercive field is a function of the measurement time, and the attempt frequency for spin flips. This measurement time can be related to the SR, and the flip frequency can be assumed to be proportional to the damping. Therefore, loops calculated with the same value of SR$/\lambda_s$ would have the same coercive field which gives the loop width. Based on this, Behbahani {\it et al.} were able to use high SR in their simulations to obtain hysteresis loops that closely match those obtained at very low SR, by employing extremely high damping values, and large time steps. Recently, this methodology was successfully applied to simulate hysteresis of iron oxide magnetic nanoparticles with application to
hyperthermia~\cite{behbahani2021multiscale}.
 
In Figure~\ref{fig:damping} we use a field frequency $f_0$ for the case with $\phi=90^{\circ}$. For $\lambda_s=0.5$ the hysteresis loop in (a) is reasonably close to the SW prediction, and the magnetization versus time in (b) indicates reasonable convergence. However, for $\lambda_s=0.1$, the loop in (a) is poorly defined, as expected from the lack of convergence towards stable magnetization values shown at the top of panel (b). The possibility of reducing the simulation time in micromagnetic simulations by increasing both the SR and the damping, leaving constant the ratio SR$/ \lambda_s$, would be equivalent, in our case, to keeping constant the ratio $f/\lambda_s$. However, SLD simulations at high damping might depart from the desired dynamics and should be treated with care. In our simulations, by choosing $\lambda_s =0.5$, we were able to set $t_{\rm sim}=90$ ps and use $f_0/4$ as discussed at the beginning of this section. For lower damping values, smaller frequencies would be required, significantly increasing the computational costs. Following the previous scaling, for instance, for $\lambda_s=0.1$, around 400 ps ($4 \times 10^6$ steps) are required to stabilize the magnetization at each field value, in contrast to the 90 ps ($9 \times 10^5$ steps) that are required for $\lambda_s=0.5$. We have run some simulations for such a low damping and nanosecond steps, and verified that this is indeed the case, explaining the lack of convergence in Figure~\ref{fig:damping}(a).


\begin{figure}[htb]
\includegraphics[width=0.99\columnwidth]{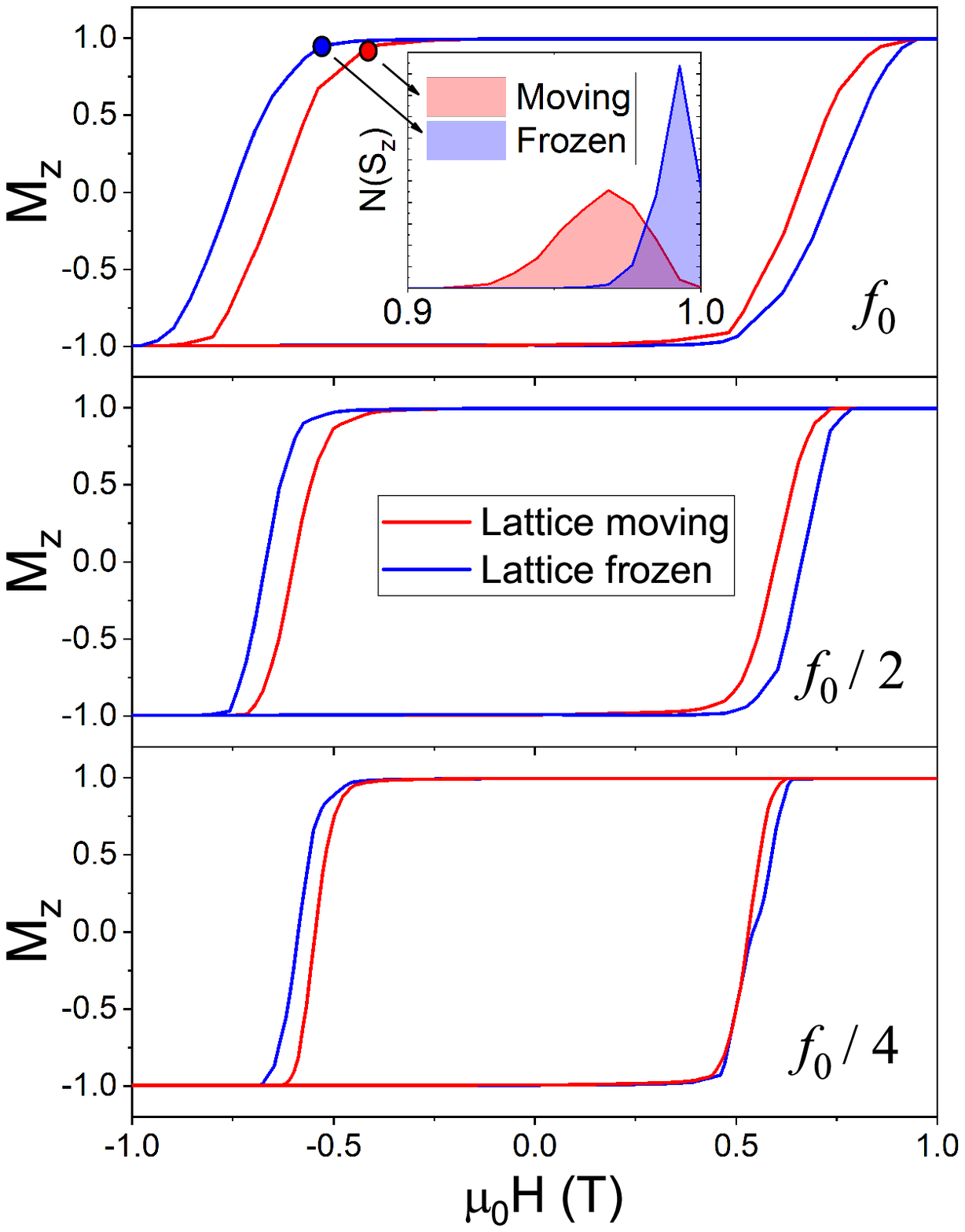}
  \caption{(Color online.) Comparison of the hysteresis loops obtained with (lattice moving) and without (lattice frozen) coupling to the lattice dynamics at $T=10 K$. All the figures correspond the case $\phi=0^{\circ}$ at different field frequencies, $f_0$ (top), $f_0/2$ (middle) and $f_0/4$ (bottom). The inset in the top panel shows histograms of the spins orientation along the field direction $z$ and correspond to the state of the system just before the beginning of the magnetization reversal marked with circles in the respective curves. 
  }
  \label{fig:lattice_frozen_moving}
\end{figure}

\subsection{Coupling the lattice dynamics}

All the results presented so far correspond to frozen-lattice simulations, meaning that the lattice vibrations are not included in the dynamics. We now analyze the effect of full SLD calculations actually coupling the spin and lattice degrees of freedom, which we refer to as moving-lattice simulations. \textcolor{black}{We analyze this effect for two different systems, perfect bcc bulk samples and nanoparticle samples including defects.}

\subsubsection{\textcolor{black}{Low temperature bulk simulations}}

 In Figure~\ref{fig:lattice_frozen_moving} we compare the simulated hysteresis loops at T=10 K for $\phi=0^{\circ}$ obtained with (moving-lattice) and without (frozen-lattice)  spin-lattice coupling at different frequencies. The former gives lower coercivity and narrower loops in comparison with the frozen-lattice case. This effect is more pronounced for higher frequencies. This is consistent with the fact that in a system that incorporates new degrees of freedom (in this case those of the lattice), the energy barrier that must be overcome to reverse the magnetization should decrease. In other words, when a reversal field is applied, a moving lattice provides additional paths for the magnetization relaxation process. However, this effect is small at a frequency $f_0/4$ where a stable magnetization is easily achieved for the chosen sweep rate, showing that the lattice dynamics contributions are not very important at these low temperatures ($T=10$ K). As it was previously shown for iron nanoparticles at equilibrium, the lattice fluctuations would become significant at temperatures higher than $300\,$K \cite{dosSantos2020, meyer2022influence}. The equilibrium magnetization for a moving lattice is always smaller than that of a frozen lattice for all temperatures \cite{meyer2022influence}.
In addition, in the present approach, the anisotropy contributions do not depend on the atomic positions nor the lattice temperature. Larger differences between moving and frozen lattice simulations might be observed for Hamiltonians that include additional terms which are sensitive to factors like atomic volume  \cite{perera2014combined, nieves2021neel}, or “effective" local electronic density \cite{caturla-dednam2022spin}.

\subsubsection{\textcolor{black}{Intermediate and high temperatures bulk simulations}}

In this subsection, we focus on intermediate temperatures, above $ \sim 1/3$ of the Curie temperature and also high temperatures, close to $T_C$. Lattice-induced spin fluctuations enter in this model through the interatomic distance dependence of the exchange function $J(r_{ij})$ in the magnetic Hamiltonian Eq.~(\ref{hamiltonian_mag}). At low temperatures, changes in $J(r_{ij})$ due to lattice vibrations are negligible. Therefore, the coupled spin-lattice dynamics show little influence in hysteresis loops in that regime, as it was shown in Figure~\ref{fig:lattice_frozen_moving} and discussed in the previous subsection.
To address the effect of the coupled spin-lattice dynamics at higher temperatures, we have run simulations for the case of uniaxial anisotropy with the easy axis aligned with the external field, $\phi=0^{\circ}$, at several temperatures, in the range of $T=300-1000$ K. For the parametrization of $J(r_{ij})$ used in this work, $T_C$ is around  $1050$ K, as discussed in Section~\ref{sec:framework}. Given the large fluctuations observed at these temperatures, we consider for these simulations a larger bulk system, including around 65000 atoms, as mentioned in the Methods section, in order to avoid finite-size effects, and use longer simulation times (180 ps per point) in order to avoid the undesirable regime in which the magnetization
cannot equilibrate to the field direction before the next field increment.  

\begin{figure}[htb]
\includegraphics[width=0.99\columnwidth]{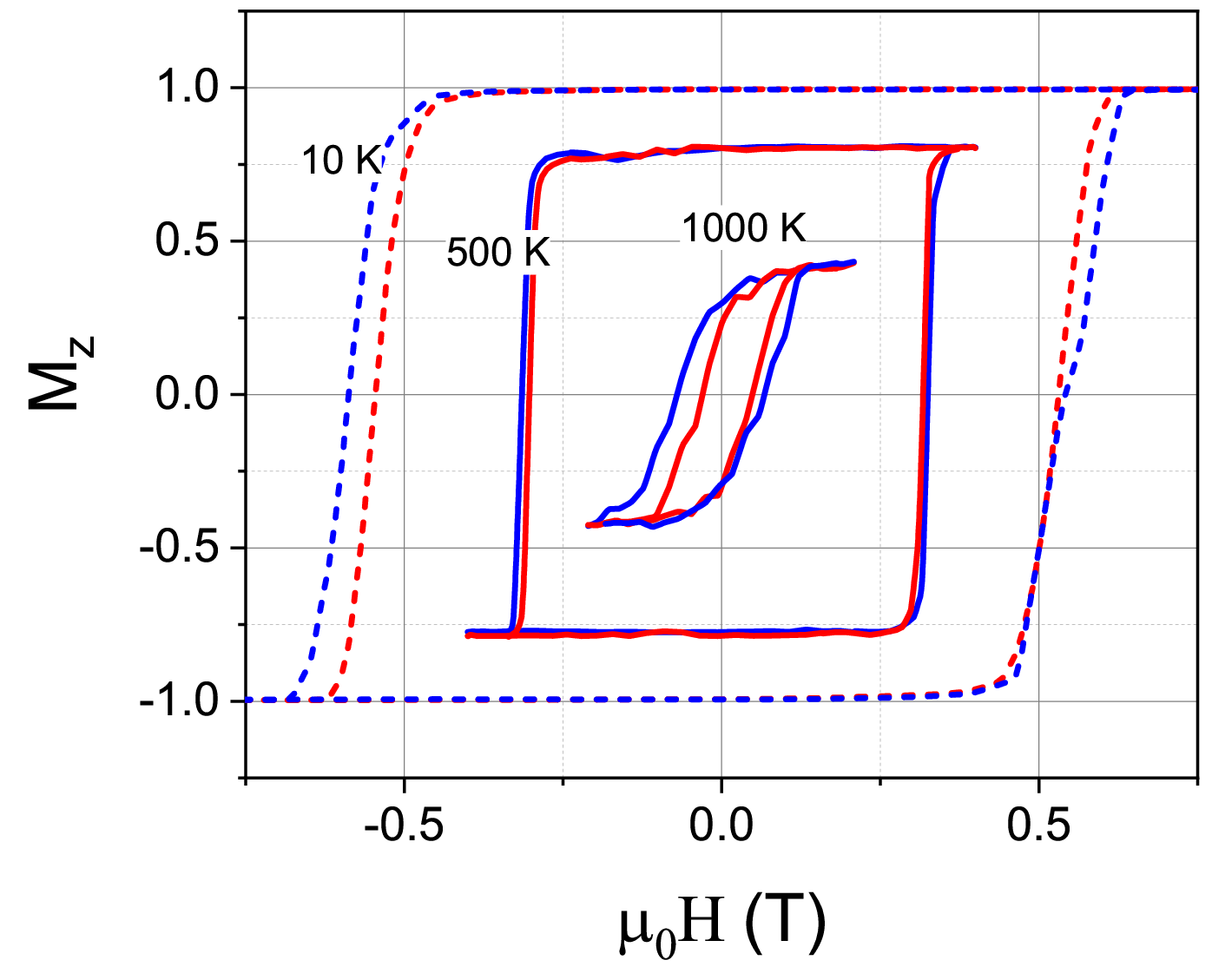}
  \caption{(Color online.) Comparison of the hysteresis loops at $T=10\,$K, $T=500\,$K and $T=1000\,$K for lattice frozen and lattice moving procedures. These results correspond to the case $\phi=0^{\circ}$. For the simulations at $T=500, 1000\,$K   a smaller sweep rate, $SR=0.5 \times 10^8$ ${\rm T}/{\rm s}$ (which corresponds to 180 ps of simulation time per point) was used. } \label{fig:lattice_frozen_moving_10K_and_500K}
\end{figure}

In Figure \ref{fig:lattice_frozen_moving_10K_and_500K}, we compare the hysteresis loops obtained at these conditions under the two different simulation approaches, lattice-frozen and lattice-moving, for $T=500\,$K and $T=1000\,$K  contrasted to the previous results at $T=10\,$K. A small difference, consistent with the results at $10\,$K and with the arguments outlined in the previous subsection, can be observed between the lattice frozen and lattice moving loops at $T=500\,$K. However, this difference is within the margin of error given by the standard deviation of the average by which the curves are obtained. This is because, at these relatively high temperatures, the effect of increased lattice vibrations on exchange and spin fluctuations is small compared to the large fluctuations caused by the thermal noise in the stochastic field $\bm{\zeta}(t)$ in Eq.(\ref{EOM_spin2}). 
This is confirmed by the spin histograms in the inset of Figure \ref{fig:lattice_frozen_moving_500K} in the supplementary material, where a very similar dispersion of spin orientations is evidenced for both frozen and moving lattice cases. 
A more pronounced difference between the frozen and moving lattice protocols is observed for the loops calculated at $T=1000\,$K. For the moving lattice simulations, the Curie temperature is close to $1050\,$K (as it was mentioned above), while for the  frozen lattice ones $T_C \sim 1150\,$K \cite{meyer2022influence}. Consequently, in the moving lattice procedure, the system is closer to a state where hysteresis should vanish and, therefore, present loops with smaller area than those of the lattice frozen simulations.

Despite the small differences obtained between the lattice moving and lattice frozen approaches, the effect of temperature is well captured by the present simulation method and protocol. At higher temperatures, smaller loops are obtained, i.e., smaller coercivity and saturation magnetization as the temperature is increased, following the expected behavior.

We note that the SW model is sometimes applied at finite temperatures, despite its lack of validity under those conditions \cite{wang2012temperature}, by re-scaling the saturation magnetization to match values at the desired temperature. If the same procedure was applied to our case, there would be strong disagreement between the model and the simulations, which represents a warning for such future comparison between the SW model and experiments at finite temperature.


\subsubsection{\textcolor{black}{Nanoparticles simulations}}

\textcolor{black}{Magnetic nanoparticles are of technological interest, and their nanostructure will affect magnetization. For example, the microstrain state inside nanoparticles can greatly modify their magnetic properties \cite{muzzi2022hardening}, including cases where lattice deformation can be driven by vacancies \cite{lappas2019vacancy}. As another example of defects influencing magnetization, experiments suggest that defects might play a role in Fe NPs remaining blocked, as opposed to defect-free NP which would behave superparamagnetically \cite{balan2014direct, kleibert2017direct}.} 

\textcolor{black}{Most atomistic spin dynamics simulations consider a frozen perfect lattice, even for cases with defects, like NP surfaces \cite{evans2014atomistic, evans2020atomistic2, hovorka2012curie, ellis2015switching}, interfaces \cite{nedelkoski2017origin, chureemart2017hybrid}, and point defects \cite{lappas2019vacancy}. Such approach allows for the use of discrete exchange values for 1st, 2nd, etc. nearest neighbors, greatly saving computational time. There are isolated efforts to include lattice distortions for a frozen lattice, which require the capability of handling magnetic interactions at arbitrary distances and not just regular lattice positions. Spatial configurations from molecular dynamics can be used as input \cite{westmoreland2018multiscale}, or set with other criteria, like in the recent study of a single dislocation dipole in bulk Cr \cite{bienvenu2023interplay}.
Sometimes, defects have been emulated by more symmetric configurations, like the the simulation of vacancies by weakly coupled ferromagnetic spin pairs, without considering lattice strain \cite{lappas2019vacancy}. In addition, these codes will not allow for dynamical lattice effects, including evolution of defects with time due to temperature, stress, and coupling to magnetic degrees of freedom. For a NP, surface roughness might vary considerably with temperature.}

\begin{figure*}[ht]
\centering
\includegraphics[width=0.99\textwidth]{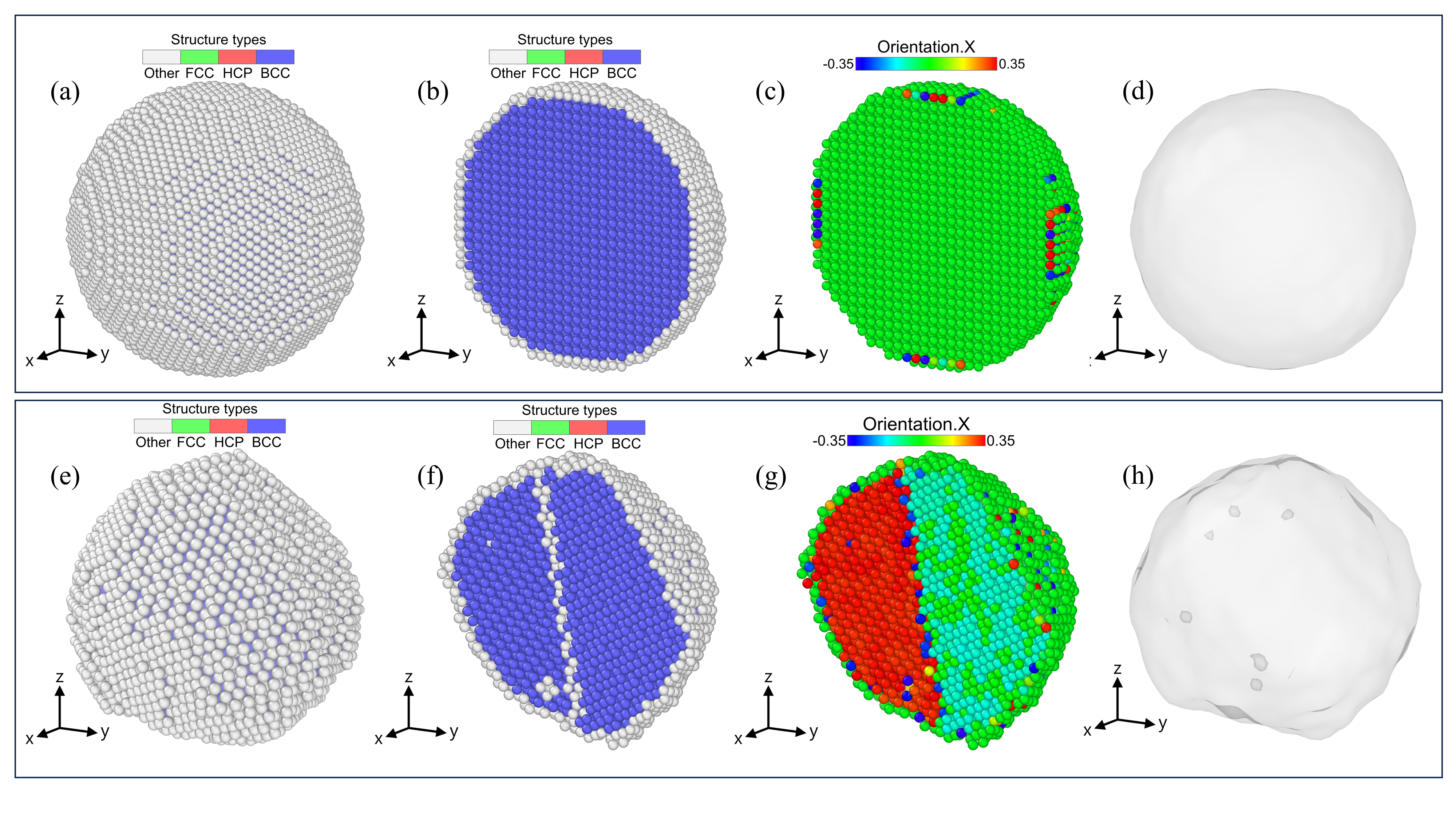}
  \caption{\textcolor{black}{(Color online) Snapshots of Fe nanoparticles employed in the hysteresis loop simulations. The upper panel ((a) - (d)) shows snapshots of the pristine NP (8 nm diameter spherical NP) while the lower panel ((e) - (h)) shows the corresponding snapshots for the defective NP. In (a) and (e) the entire NP is shown, with atoms colored by PTM structure type. Surface atoms are classified as “other” (unknown structure). In (b) and (f) a slice near the center of the NP is shown, also colored by PTM structure type. In (f), atoms in a twin boundary crossing the NP, and around a vacancy are also classified as “other”. Figures (c) and (g) show the same slice as (b) and (f), but atoms are colored according to their PTM atomic orientation along the x-axis, to better show the twin rotation in the defective NP. Finally, in (d) and (h) a surface mesh is shown for the NP surface, a few single vacancies, and vacancy clusters inside the defective NP.}} \label{fig:NP_snapshot_pane}
\end{figure*}

\begin{figure}[htb]
\includegraphics[width=0.99\columnwidth]{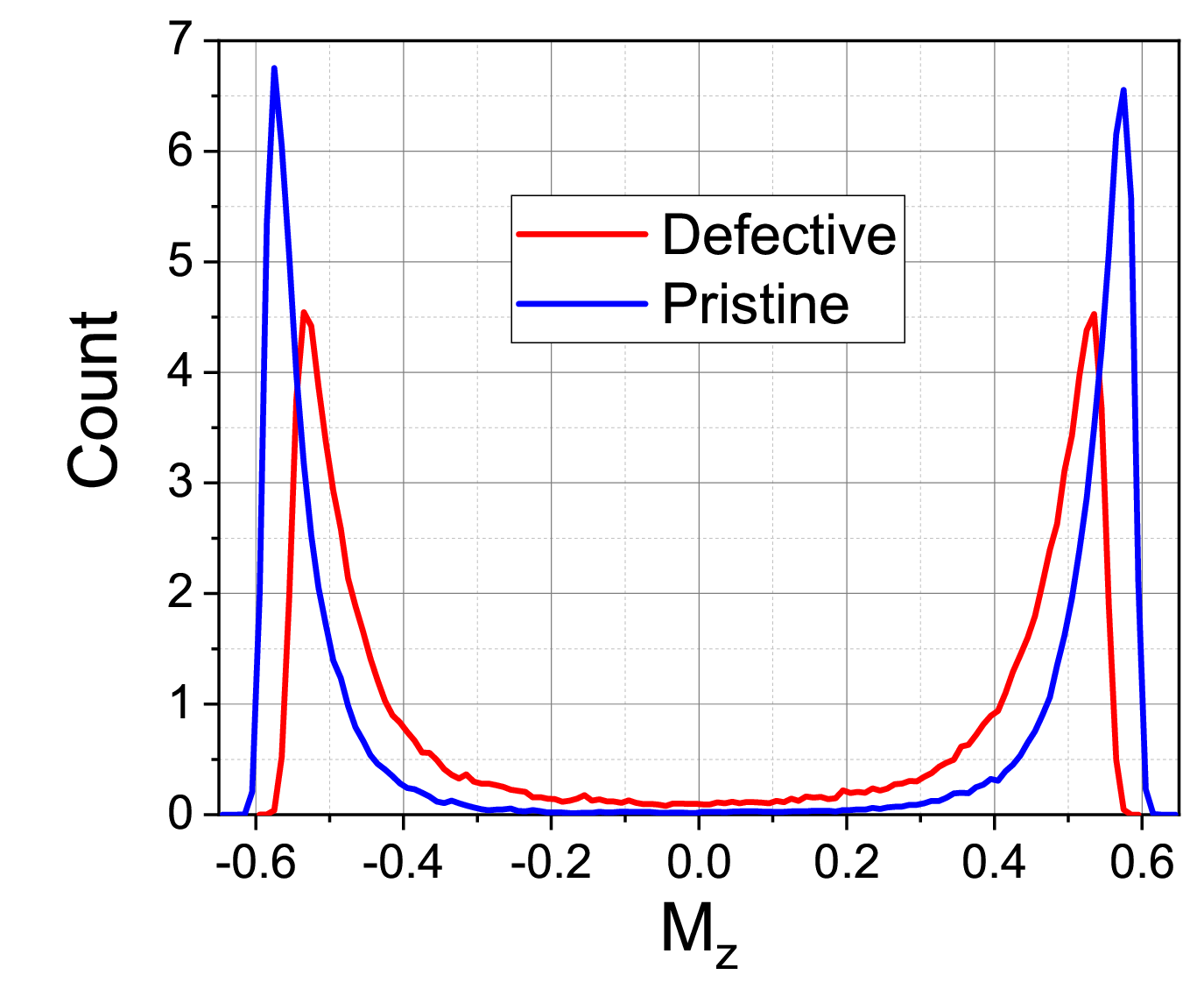}
  \caption{\textcolor{black}{(Color online.) Histograms of the magnetization along the field direction ($M_z$), for the defective and the pristine NPs. The histograms were built with the data obtained directly from the entire simulation and are normalized to unit area.} }\label{fig:Histograms_NP}
\end{figure}

\begin{figure}[htb]
\includegraphics[width=0.99\columnwidth]{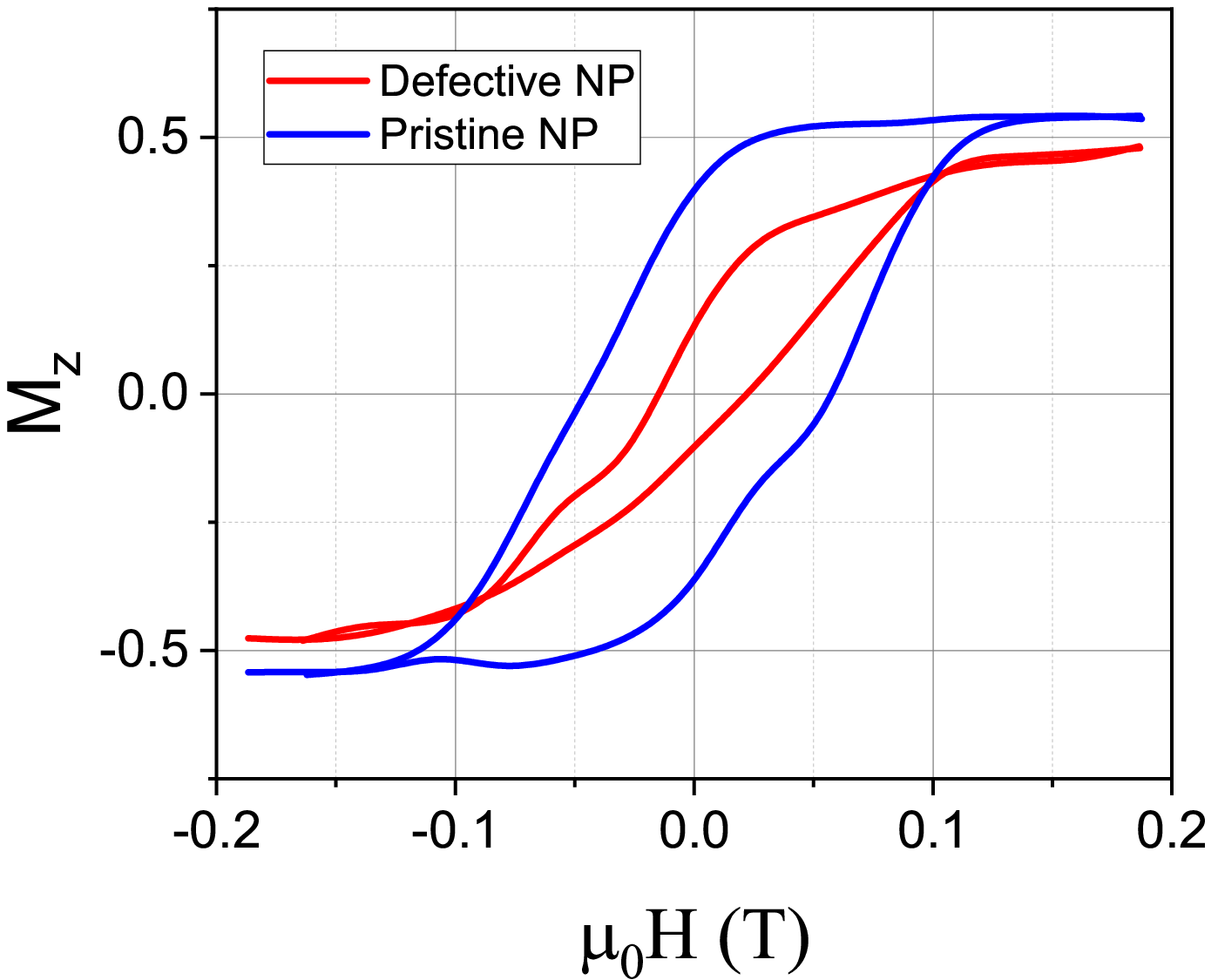}
  \caption{\textcolor{black}{(Color online.) Hysteresis loops comparing defective and pristine nanoparticles (NPs). The hysteresis loop area of the defective NP is nearly 4 times smaller than that of the pristine NP.
}} \label{fig:Loops_NP}
\end{figure}

\textcolor{black}{In order to explore the role of lattice dynamics, we simulate hysteresis loops for pristine and defective Fe NPs samples. In principle, one could also compare the magnetization of the defective NP with a frozen lattice with exactly the same NP with a moving lattice. Here we focus on the extreme comparison of a pristine NP with frozen lattice and a defective NP with moving lattice, i.e coupling the spin and lattice dynamics, given that we would like to emphasize the potential of the method compared to the current standard simulations that generally lack the inclusion of MD simulations configurations and employ a fixed, ideal lattice structure. Figure \ref{fig:NP_snapshot_pane} shows the “perfect” NP used for the frozen lattice simulations, alongside the Fe NP with defects. It can be seen that the defective NP has a non-spherical topology, with large planar facets, and also includes a twin boundary, and a few vacancy clusters. Figure \ref{fig:Histograms_NP} shows the histograms of magnetization values over the entire simulation, covering several hysteresis cycles for both NPs. The differences between the two NPs is large, with the defective NP experiencing significantly larger and more frequent magnetization fluctuations. This suggests that, for a NP, defects which can be included in spin-lattice dynamics display different qualitative and quantitative behavior than the absence of defects. The larger fluctuations lead to narrower hysteresis loops, as qualitatively expected. Figure \ref{fig:Loops_NP} shows hysteresis loops for both NPs, with the defective NP displaying lower saturation magnetization and lower coercive field, resulting in a significantly lower loop area, approximately one-quarter that of the perfect NP. We obtain a coercive field around 50 mT for the pristine frozen NP and around 18 mT for the defective NP at 500 K. This is for a single, oriented NP, with a diameter of 8 nm. Loops from other experiments \cite{balan2014direct} show smaller coercive fields, around 3 mT. However, our value for the our defective NP shows very good agreement with an experimental value of around 20 mT for a collection of Fe NP of the same size as in MD, at 300 K \cite{ibusuki2001magnetic}. Furthermore, the experimental coercive field is approximately half of what we have obtained for the pristine NP. Given that the coercive field of a random collection of NP is expected to result in a coercive field of about half that for oriented NP \cite{cullity2011introduction}, the agreement is reasonable. In addition, the hysteresis loop of a prismatic Fe NP ($210\times210\times15$ $nm^3$) gives $H_c \sim 75$ mT for the field at $45^{\circ}$, according to recent micromagnetic simulations \cite{sudsom2020micromagnetic}, also indicating that our simulations obtain reasonable coercive field values. 
For experimental measurement times of seconds, NP below some critical size are expected to behave superparamagnetically at high temperatures \cite{cullity2011introduction,hedayatnasab2017review}, but for our NP size and simulation times of ns, the NPs behave ferromagnetically. Recent experiments observed that, within a collection of Fe NP with 5-20 nm diameter, many behaved with the expected superparamagnetism, but some remained in a blocked ferromagnetic state \cite{balan2014direct,kleibert2017direct}. Defects were assumed to be the cause for the ferromagnetic behavior, opposite to what we find in our simulations, where the defects considered here facilitate magnetization flips and decrease the coercive field. These preliminary simulations of pristine and defective NPs highlight the importance of SLD calculations. The introduction of defects, and the possibility that they can evolve and migrate during the simulation time, can change the magnetic behavior of such samples.}

\section{\label{sec:discussion} Discussion}
Given the short timestep required to integrate the atomic degrees of freedom, usually $\sim 0.1$ fs - 1 fs, one needs relatively long simulations to achieve a stable magnetization at a given value of the applied magnetic field. Typically, this would involve several precession periods. Using the Larmor frequency for Fe gives a period of $\sim 36\,$ps for a field of $1\, $T. In order to speed-up the spin-dynamics convergence, a Gilbert damping $\lambda_s=1.0$ is used in most ASD simulations \cite{bottcher2011atomistic,westmoreland2020atomistic,alzate2019optimal,jenkins2021atomistic} and values much larger than 1 have been used in micromagnetic simulations of hysteresis loops ~\cite{behbahani2020coarse}.  In our simulations, we found that for a given damping, as frequency decreases, the loop area also decreases until reaching a constant value. For bulk iron, and a Gilbert damping of $0.5$, we found that frequencies of $125$ MHz or lower provide such a constant loop area, for simulation times of $90$ ps for each field. Values between 30 and 200 ps (depending on the system under study and the Gilbert damping employed) are also used for ASD simulations \cite{bottcher2011atomistic, alzate2019optimal,jenkins2021atomistic}. For the case in which the anisotropy axis and the field direction are perpendicular, $\phi=90^{\circ}$, a converged loop can be found at even higher frequencies. Agreement between SLD and SW suggests that loops using this limiting frequency would not be much different from the experimental or theoretical estimates. This result, partially solves the problem related to the computational costs in atomistic simulations of hysteresis loops, allowing to reproduce experimental hysteresis loops, running SLD simulations at much higher frequencies than those typically used in experiments. Large computer clusters allow for simulations of increasingly larger systems by distributing spatial scales, but time scale cannot be split amongst different parallel processes. Therefore, new approaches for accelerated dynamics \cite{plimpton-voter2020acceleratedDynamics}, bypassing this time scale problem, will be required for more efficient simulations of hysteresis loops with SLD. 

When atomic motion is coupled to spin dynamics we found that, for the bulk simulations, the lattice dynamics contributions depend on the loop frequency, but they are typically small at low and intermediate temperatures, but more notorious at higher temperatures close to $T_C$. At low temperatures ($10\,$~K) atoms vibrations are negligible. At intermediate temperatures ($300-500\,$K), fluctuations in spin-spin exchange $J(r_{ij})$ produced by the atomic motion  are outweighed by the thermal fluctuations induced by the stochastic term in the sLL equation of the spin dynamics. \textcolor{black}{ In the case of the NP simulations we find significant differences, both qualitative and quantitative, between the hysteresis loops of a pristine NP with a frozen lattice and the loops foor a defective NP with a moving lattice, both with a spin temperature of $T=500$ K.}

Although the effects of the coupled spin-lattice dynamics are small for bulk bcc iron, in more complex systems that exceed the scope of this work, there might be dynamical effects that would result in different hysteresis loops if the lattice and spin dynamics are coupled. Also, future studies could include a quantum thermostat to improve the description of thermodynamic properties at low temperatures \cite{woo2015quantum}. \textcolor{black}{Equations of motion which would provide angular momentum conservation might be explored too \cite{caturla-dednam2022spin,cooke2023angular}}.

SLD simulations could be applied to hysteresis loops for other systems of technological interest, such as more complex nanoparticles \cite{lak2021embracing}, \textcolor{black}{including core-shell nanoparticles with realistic interfaces \cite{oberdick2018spin}}, magnetostrictive materials with magnetic microstructures \cite{he2018determination}, compounds like CrN, where the spin-lattice coupling is responsible for the unusual temperature dependence of the thermal conductivity \cite{stockem2018anomalous}, and bulk systems with defects \cite{palanisamy2021influence, han2021ultrastrong,bersweiler2021revealing,jenkins2021atomistic,meyer2022influence}. Further examples include strained antiferromagnetic materials that exhibit different responses to external fields \cite{van2022strain} and other materials where strain affects the magnetic properties~\cite{li2022strain,shen2022enhanced}. Finally, recent theoretical studies suggest topological magnon phase transitions tuned by time-dependent strains in 2D materials \cite{Vidal_PhysRevB.106.224401}. In binary alloys like Fe-Cr the formation of stable precipitates is known to occur at certain concentrations, and this would dynamically affect magnetic properties \cite{erhart2008short}, as well as magnetically driven phase transformation \cite{niu2018magnetically}.  Magnetic alloys can display complex microstructure, including dislocations, stacking faults, twins, etc., which greatly affects their magnetic properties \cite{zhao2021microstructural}. In particular, High Entropy Alloys (HEA) are materials of technological interest, with a complex microstructure, including dislocations, twins, precipitates, and different crystallographic phases, which leads to a complex magnetic behavior \cite{kumari2022comprehensive}, pressure-induced phase stabilities, and magnetovolume effects \cite{liu2019pressure}. 
For all of the above cases, SLD could provide information which can inform micromagnetic and atomistic spin dynamics simulations. 

\section{\label{sec:conclusions} Summary and Conclusions}
In this study, we apply a systematic approach toward incorporating Spin-Lattice Dynamics in the simulation of hysteresis phenomena. Each step of the process was thoroughly evaluated and optimized to ensure the accuracy and reliability of the simulation results. We used SLD simulations to calculate the hysteresis loops of a bulk ferromagnetic system at different temperatures \textcolor{black}{and the hysteresis loops of bcc Fe NPs with the inclusion of structural defects.} SLD can include temperature effects (thermal spin fluctuations as well as lattice vibrations) and lattice defects in a relatively simple way, complementing spin-dynamic or micromagnetic simulations.

\textcolor{black}{We have also performed formal calculations to derive the Fokker-Planck equation associated to the set of coupled Langevin equations Eqs.(\ref{EOM-p1})-(\ref{EOM_spin2}). From this, we deduce the correct expressions for the amplitudes of the white noise correlations that allow the system to reach an asymptotic stationary Gibbs-Boltzmann equilibrium state.} 

From our simulations at a temperature of $10 \,$K, spins behave almost like a macrospin, but there is some spread in the magnetization values, associated with thermal effects and exchange interactions. The exchange energy, which is much larger than 
the Zeeman or anisotropy energies, does not significantly affect the hysteresis loops at low temperatures. The coercive field depends linearly on the magnitude of the uniaxial anisotropy, as in the SW model. Cubic anisotropy and uniaxial anisotropy produce nearly the same loops for the external field aligned with an easy anisotropy axis. However, there are important differences for the case of $\phi \approx 54.7^{\circ}$ misalignment, for which the applied field lies along the [111] direction. For this case, cubic anisotropy leads to a lower coercivity value. 

In order to validate our low-temperature SLD results, we compared to the Stoner-Wohlfarth model, which is widely used in experimental and simulation studies. In the SW model, there is no lattice, no temperature, and no exchange interactions. The model assumes uniaxial anisotropy and Zeeman energy contributions, and spins in the volume are assumed to behave like a single macrospin. We compared the outcome of the SLD simulations at $T=10$ K to the predictions of the SW model for different angles between the external field and the anisotropy axis. We analyzed the effects of several parameters like simulation run time, field frequency, damping, exchange, anisotropy and lattice-vibrations. We found that the low-temperature loops obtained with SLD agree very well with those predicted by the SW model, provided that several parameters are chosen with care. Once this agreement is established, validating our model, we study the effect of the coupled spin-lattice dynamics on the intermediate and high temperature hysteresis loops where the SW model is no longer valid. 

\textcolor{black}{Our study employs several approximations, including a classical Heisenberg Hamiltonian, but it can capture reasonably well the behavior of lattice defects that modify magnetic properties. In particular, we are able to model the high-temperature evolution of a nanoparticle with both point and extended defects, where such defects lower the overall magnetization and significantly reduce the area of the hysteresis loop, compared to a typical frozen lattice approximation. This was qualitatively expected, given that disorder hinders global magnetization and, under varying external fields, helps fluctuations which would reduce loop area. However, quantitative simulations can certainly contribute to the understanding of experimental results, and with the future design of tailored magnetization in NP.}

In order to model complex materials of interest, it is desirable to have reliable simulation techniques and protocols for hysteresis processes that incorporate the effects of coupled lattice and spin dynamics and this study would be a contribution in that direction. While most of our results correspond to the already well-studied case of pure bulk Fe, our approach provides a framework for future studies to efficiently use SLD for the study of hysteresis behavior in magnetic materials in general. 

\begin{acknowledgments}
We thank Romina Aparicio for her valuable contribution 
through support simulations and helpful discussions. GDS 
and EMB thank support from a SIIP-UNCUYO-2022-2023 grant, 
from PICTO-UUMM-2019-00048 and from PIP 2021-2023 11220200102578CO. We thank Sergei Dudarev for sharing his valuable notes on spin Langevin damping. We thank computer run time in the cluster TOKO (toko.uncu.edu.ar). The authors thank support from IPAC-2019 grant from Sistema Nacional de Computación de Alto Desempeño (SNCAD), for run time in the cluster Dirac http://dirac.df.uba.ar/. FR acknowledges financial support 
from CONICET (Argentina)  under Project No. PIP 112-202001-01294-CO and Universidad Nacional de San Luis (Argentina) under Project No. PROICO 03-2220. SC benefited from an Erasmus+ Grant during his stay at LPTHE, Paris.
\end{acknowledgments}

\appendix*
\section{}
The aim of this Appendix is to derive the Fokker-Planck equation associated to the set of coupled Langevin equations, Eqs.(\ref{EOM-p1})-(\ref{EOM_spin2}), which rule the translational and rotational dynamics of the magnetic moments. 
From the Fokker-Planck equation, we then deduce the amplitudes of the white noise correlations, Eqs.(\ref{DL}) and (\ref{DS}), that allow the system to reach an asymptotic stationary state of equilibrium Gibbs-Boltzmann form. 

For the sake of completeness, we work with a Langevin equation for spins in which we add the stochastic field to
both the gyromagnetic and the relaxation terms, with a parameter $A$ controlling the presence or absence of noise in the latter dissipative mechanism.  In this way, for $A=1$ we obtain the sLLG equation while for $A=0$ we recover our sLL Eq.({\ref{EOM_spin2}}) \cite{garcia1998langevin}.

\subsection{Langevin formulation}

We take a set of $i=1,\dots, N$ particles at positions $\bm{r}_i = (r^i_x,r^i_y,r^i_z)$, with momenta
$\bm{p}_i = (p^i_x,p^i_y,p^i_z)$ and magnetic moments $\bm{s}_i = (s^i_x,s^i_y,s^i_z)$.   As in the main text, 
we label $a$, $b$, or $c$ the space coordinates $x$, $y$, and $z$. Since the number of components of the 
magnetic moments are also three we use the labels $a,b,c$ for them as well. 
The magnetic moments are normalised such that $|\bm{s}_i|^2 = (s^i_a)^2 = 1$ for each particle~$i$.
Here and in the following we use Einstein summation notation over repeated $a, b, c$ indices; 
not over $i,j$ particle indices, and we write the sums over these indices explicitly when needed.

The stochastic Langevin equations of motion Eqs.(\ref{EOM-p1})-(\ref{EOM_spin2}) are recast in the generic form \cite{Aron14}
\begin{eqnarray}
d_t r^i_a &=& \frac{p^i_a}{m^i}  \; ,   \label{eq:tranchida-langevin1} \\
d_t p^i_a &=& \sum_{j(\neq i)}^N\left[-\frac{\partial V(r_{ij})}{\partial r^{i}_a} + \frac{\partial J(r_{ij})}{\partial r^{i}_a} \bm{s}_i \cdot \bm{s}_j \right] \nonumber \\
&&-\frac{\gamma_L}{m^i} p_a^i + \xi_a^i  \; ,
\label{eq:tranchida-langevin2}
\\
{d_t s}^i_a &=& \; g_{ab}^i\; \omega^i_b  \; + \overline g^i_{ab} \; \zeta^i_b 
\; , 
\label{eq:tranchida-langevin3}
\end{eqnarray}
where $d_t$ denotes the time derivative $d/dt$ and $\omega^i_b=- \frac{1}{\hbar} \frac{\partial \mathcal{H}_{\mathrm{mag}}}{\partial s^i_b}$  is the component $b$ of the effective field acting on spin $i$.  The noises affecting the translation and magnetic degrees of freedom, $ \xi^i_a$ and $ \zeta^i_a$, are independent Gaussian white random variables with zero mean and correlations
\begin{equation}
\langle \xi_a^i(t) \xi_b^j(t') \rangle = 2 D_L \delta_{ab}\delta^{ij}\delta(t-t')
\; ,
\end{equation}
and 
\begin{equation}
\langle \zeta^i_a(t) \zeta^j_b(t') \rangle = 2D_s \delta_{ab} \delta^{ij} \delta(t-t')
\; . 
\end{equation}
We will fix the coefficients $D_L$ and $D_s$ below.
The matrices $g_{ab}^i$ and $\overline g^i_{ab}$ are defined as   
\begin{eqnarray}
g^i_{ab} &=& \frac{1}{1+\lambda_s^2} \; [\epsilon_{abc}s^i_c + \frac{\lambda_s}{s_i} 
(s_i^2\delta_{ab} - s^i_a s^i_b) ]
\label{eq:g-def}
\; , 
\\
{\overline g}^i_{ab} &=& f^i_{ab} + A h^i_{ab}
\; , 
\end{eqnarray}
with
\begin{eqnarray}
f^i_{ab} &=& \frac{1}{1+\lambda_s^2} \; \epsilon_{abc}s^i_c = \frac{1}{2} (g^i_{ab}-g^i_{ba})
\label{eq:f-def}
\; , 
\\
h^i_{ab} &=& \frac{\lambda_s}{s_i} \frac{1}{1+\lambda_s^2} (s_i^2\delta_{ab} - s^i_a s^i_b ) 
\nonumber\\
&=& \frac{1}{2} (g^i_{ab}+g^i_{ba})
\label{eq:h-def}
\; .
\end{eqnarray}
Here, $f$ and $h$ are the antisymmetric and symmetric parts of $g$, respectively, and $\epsilon_{abc}$ is the completely antisymmetric Levi–Civita tensor.  Note that for $A=0$ Eq.~(\ref{eq:tranchida-langevin3}) corresponds to the sLL Eq.~(\ref{EOM_spin2}).  Instead, since $f^i_{ab}+h^i_{ab} =g^i_{ab}$, for $A=1$ we have that ${\overline g}^i_{ab} = g^i_{ab}$ and it is then 
easy to show that Eq.~(\ref{eq:tranchida-langevin3}) 
reduces to the well-known sLLG equation.
Importantly enough, we have re inserted the 
modulus of the local spins, $s_i$, since although it is fixed to 
be one, its variations with respect to the various spin components 
are non-trivial and have to be taken into account in the 
calculations that follow. For instance, 
\begin{eqnarray}
\frac{\partial s_i^2}{\partial s^i_a} = 2s_a
\; , \quad\qquad
\frac{\partial s_i^2}{\partial (s^i_a)^2} = 6
\end{eqnarray}
for each spin $i$. In the second equation a sum over the three $a$ components 
was implicit and led to a factor $3$.

\subsection{The Fokker-Planck Equation}

Let us join the momenta, position and magnetic variables of all particles, each with three components, in a single $\alpha = 1, \dots, 9N$ component vector
\begin{equation}
\bm{y} = (\bm{p}_1, \dots, \bm{p}_N, \bm{r}_1, \dots, \bm{r}_N, \bm{s}_1, \dots, \bm{s}_N )
\; . 
\end{equation}

Starting from the Chapman-Kolmogorov equation 
\begin{equation}
    P(\bm{y},t+\Delta t) 
    = 
    \int d\bm{y}_0 \; P(\bm{y}, t+\Delta t | \bm{y}_0,t)P(\bm{y}_0,t) 
    \, , 
    \label{ChapKol}
\end{equation}
we obtain the Fokker-Planck description from the Langevin one by using the definition of $P(\bm{y},t)$ via the Dirac delta function, 
\begin{equation}
P(\bm{y}, t+\Delta t | \bm{y}_0,t) = \langle\delta(\bm{y} - \bm{y}_{\xi,\zeta}(t+\Delta t))\rangle \; .
\end{equation}
This compact Dirac delta notation indicates a product over all the $9N$ components 
of the vector 
${\bm y}$ which is forced to take the value given by the solution to the Langevin equations.  
The mean value $\langle \dots \rangle$ is taken over the two noise sources and $\bm{y}_{\xi,\zeta}(t+\Delta t)$ is the solution to the Langevin equations~(\ref{eq:tranchida-langevin1})-(\ref{eq:tranchida-langevin3}) evaluated at time $t+\Delta t$, with initial condition $\bm{y}(t) = \bm{y}_0$, which clearly depends on the noises. We now expand $\bm{y}_{\xi,\zeta}(t+\Delta t)$ around  $\bm{y}_0$
since the time increment $\Delta t$ is small and the $\bm{y}$ increment $\Delta \bm{y}$ as well. Using $\bm{y} = \bm{y}_0 + \Delta \bm{y}$, 
\begin{eqnarray}
P(\bm{y}, t+\Delta t | \bm{y}_0,t) &=& \delta(\bm{y} - \bm{y}_0) - \partial_\alpha \left(\delta(\bm{y} - \bm{y}_0) \langle\Delta y_\alpha \rangle \right) \nonumber \\ 
&&+ \frac{1}{2}\partial_\alpha\partial_\beta \left( \delta(\bm{y} - \bm{y}_0) \langle \Delta y_\alpha \Delta y_\beta \rangle \right) \nonumber \\
&&+\mathcal{O}(\Delta t^2) \; , \label{Expan}
\end{eqnarray}
with summation over repeated $\alpha, \beta=1, \dots, 9N$ indices.
Combining Eqs.~(\ref{ChapKol}) and (\ref{Expan}), and integrating over $\bm{y}_0$, one gets
\begin{eqnarray}
P(\bm{y}, t+\Delta t) &=& P(\bm{y}, t) - \partial_\alpha\left(\langle\Delta y_\alpha \rangle P(\bm{y}, t)\right) \nonumber \\
&&+ \frac{1}{2}\partial_\alpha\partial_\beta \left(\langle \Delta y_\alpha \Delta y_\beta \rangle P(\bm{y}, t) \right) \nonumber \\ &&+\mathcal{O}(\Delta t^2) \; . \label{eq:FP-rhs}
\end{eqnarray}
Taking the limit for the differential of $P$, 
\begin{equation}
 \partial_t P(\bm{y}, t) =\lim_{\Delta t \to 0}\frac{P(\bm{y}, t+\Delta t) - P(\bm{y},t)}{\Delta t} \; ,
\end{equation}
and eliminating any term of higher order than $\Delta t$ in the right-hand-side of Eq.~(\ref{eq:FP-rhs}), we obtain the Fokker-Planck equation 
\begin{eqnarray}
\partial_t P(\bm{y}, t) &=&
- \partial_\alpha \left[ \frac{\langle \Delta y_\alpha\rangle}{\Delta t} P(\bm{y},t) \right] \nonumber \\
&&+ \frac{1}{2} \partial_\alpha\partial_\beta \left[  \frac{\langle \Delta y_\alpha \Delta y_\beta \rangle}{\Delta t} P(\bm{y},t) \right]
\, . 
\label{eq:Fokker}
\end{eqnarray}

The next step is to calculate the averages $\langle \Delta y_\alpha\rangle$ and $\langle \Delta y_\alpha\Delta y_\beta\rangle$
to leading order in $\Delta t$ using the Langevin Eqs.~(\ref{eq:tranchida-langevin1})-(\ref{eq:tranchida-langevin3}) 
which read, in discrete time,
\begin{eqnarray}
\Delta r^i_a &\equiv& r^i_a(t+\Delta t) -r^i_a(t)   = 
\frac{p^i_a}{m^i} \; \Delta t \; , \\
\Delta p^i_a &\equiv& p^i_a(t+\Delta t) - p^i_a(t) \nonumber \\  
&=& \sum_{j(\neq i)}^N\left[-\frac{\partial V(r_{ij})}{\partial r^{i}_a}  + \frac{\partial J(r_{ij})}{\partial r^{i}_a} \bm{s}_i \cdot \bm{s}_j\right]\Delta t \nonumber \\
&&- \frac{\gamma_L}{m^i} p_a^i \, \Delta t + \xi_a^i \, \Delta t \; , \\
\Delta s^i_a &\equiv& s^i_a(t+\Delta t) - s^i_a(t)   \nonumber \\
&=&  g_{ab}^i\; \omega_b^{i} \; \Delta t \; + {\overline g}_{ab}^i \; \zeta_b^i \; \Delta t \;.   \label{eq:Langevin-discrete}
\end{eqnarray}
All variables in the right-hand-sides of these equations are evaluated at the mid-point $\bm{y}_{\rm mp} = [\bm{y}(t) + \bm{y}(t+\Delta t)]/2$ since we chose to work with the Stratonovich prescription, the unique scheme consistent with the conservation of the modulus of the magnetization, in 
the way we wrote the equations~\cite{garcia1998langevin,Aron14}. 
When working at ${\mathcal O}(\Delta t)$ we will be able to, in some cases, replace this mid-point by the initial one $\bm{y}_0$ 
in the interval $[t, t+\Delta t]$.

The averages of the position increments or the product of two position increments are
\begin{eqnarray}
\langle \Delta r_a^i \rangle &=& \langle p_a^i \frac{\Delta t}{m^i}\rangle = \frac{p_a^i}{m^i}\Delta t + \mathcal{O}(\Delta t^2)
\; , \label{inc_r} \\
\langle \Delta r^i_a \; \Delta r^j_b \rangle &=& \mathcal{O}(\Delta t^2)
\; .
\end{eqnarray}
Besides, for the momentum and magnetisation components we obtain, respectively,
\begin{eqnarray}
\langle \Delta p_a^i \rangle &=&  \sum_{j(\neq i)}^N\left[-\frac{\partial V(r_{ij})}{\partial r^{i}_a}  + \frac{\partial J(r_{ij})}{\partial r^{i}_a} \bm{s}_i \cdot \bm{s}_j \right]\Delta t \nonumber \\
&&- \frac{\gamma_L}{m^i} p_a^i \Delta t + \mathcal{O}(\Delta t^{3/2})
\; , \\
[10pt]
\langle \Delta p^i_a \; \Delta p^j_b \rangle &=& 2 D_L \delta_{ab} \delta^{ij}\Delta t + \mathcal{O}(\Delta t^{3/2})
\; ,
\nonumber\\
\end{eqnarray}
and 
\begin{eqnarray}
\langle \Delta s_a^i \rangle &=& g_{ab}^i\; \omega_b^{i} \; \Delta t -2 D_s \dfrac{1 + \left(A \lambda_s \right)^2}{(1+\lambda_s^2)^2} \; s^i_a \, \Delta t \nonumber \\
&& + \mathcal{O}(\Delta t^{3/2}) 
\; , \\
[10pt]
\langle \Delta s^i_a \; \Delta s^j_b \rangle &=& 2 D_s  \dfrac{ 1  + (A\lambda_s)^2 }{(1+\lambda_s^2)^2} \; (s^2_i \delta_{ab} - s^i_a s^i_b) 
\, \delta^{ij} \Delta t 
\nonumber \\
[10pt]
&&+\mathcal{O}(\Delta t^{3/2}) 
\; . 
\label{inc_ss}
\end{eqnarray}
The averages over the cross products are sub-leading since the noises $\xi^i_a$ and $\zeta^i_a$ are not correlated.
Replacing now Eqs.~(\ref{inc_r})-(\ref{inc_ss}) into Eq.(\ref{eq:Fokker}), we finally find the explicit Fokker-Planck equation
\begin{widetext}
\begin{eqnarray}
 \partial_t P &=&
- \frac{\partial}{\partial r^i_a} 
\left( \frac{p_a^i}{m^i}  \, P \right)
- \frac{\partial}{\partial p^i_a} 
\left\{ 
\sum_{j(\neq i)}^N
\left[-\frac{\partial V(r_{ij})}{\partial r^{i}_a}  + \frac{\partial J(r_{ij})}{\partial r^{i}_a} \bm{s}_i \cdot \bm{s}_j \right]\, P - \frac{\gamma_L}{m^i} p_a^i \, P \right\}
+ \frac{1}{2}
\frac{\partial}{\partial p^i_a} \frac{\partial}{\partial p^i_b} 
\left( 
2 D_L \delta_{ab}  \, P
\right)
\nonumber\\
&&
- \frac{\partial}{\partial s^i_a} 
\left[g^i_{ab} \; \omega_b^{i} \; P - 2 D_s \frac{1+(A\lambda_s)^2}{(1+\lambda_s^2)^2} s^i_a \;P \right]
+ \frac{1}{2} \frac{\partial}{\partial s^i_a} \frac{\partial}{\partial s^i_b} 
\left[2D_s \frac{1+(A\lambda_s)^2}{(1+\lambda_s^2)^2}\left( s^2_i \delta_{ab} - s_a^i s_b^i\right) P \right]
\; , 
\label{eq:FP}
\end{eqnarray}

where we recall that there is a sum over repeated $a$ 
and $b$ indices, and here also the $i$ index.
\end{widetext}

\begin{widetext}
\subsection{Stationary solution}

We now check whether the Gibbs-Boltzmann  distribution $P_{\mathrm{eq}} = \mathcal{Z}^{-1}e^{-\mathcal{H}/k_B T}$, where $\mathcal{Z}$ is the partition function, is a stationary solution of the Fokker-Planck Eq.~(\ref{eq:FP}). Such a stationary $P$ does not depend on time and should satisfy $\partial_t P =0$, {\it i.~e.}

\begin{eqnarray}
 0 &=& 
 - \sum_i^N\frac{p_a^i}{m^i} \,\frac{\partial P}{\partial r^i_a}
- \sum_{ i\neq j }^N\left[-\frac{\partial V(r_{ij})}{\partial r^{i}_a}  + \frac{\partial J(r_{ij})}{\partial r^{i}_a} \bm{s}_i \cdot \bm{s}_j \right]\, \frac{\partial P}{\partial p^i_a} + 3\sum_i^N\frac{\gamma_L}{m^i}\, P + \sum_i^N\frac{\gamma_L}{m^i} p_a^i \frac{\partial P}{\partial p^i_a} + D_L
\sum_i^N \frac{\partial ^2 P}{\partial (p^{i}_a)^2} 
\nonumber\\
&& + 2 \frac{\lambda_s}{s_i} \frac{1}{(1+\lambda_s^2)} \sum_i^N s^i_b \; \omega_b^{i} \; P
-  \sum_i^N g^i_{ab} \; \omega_b^{i} \; \frac{\partial P}{\partial s^i_a} 
+ 6D_s N \; \frac{1+(A\lambda_s)^2}{(1+\lambda_s^2)^2}\;P
+ 2 D_s \; \frac{1+(A\lambda_s)^2}{(1+\lambda_s^2)^2}  \sum_i^N s^i_a\;\frac{\partial P}{\partial s_a^i}
\nonumber\\
&&
+D_s \frac{1+(A\lambda_s)^2}{(1+\lambda_s^2)^2}\left[-6N P - 4 \sum_i^N s_a^i \frac{\partial P}{\partial s_a^i} +  \sum_i^N  \left(s^2_i \delta_{ab} - s_a^i s_b^i\right)\frac{\partial^2 P}{\partial s_a^i \partial s_b^i}\right] 
\; ,
\label{eq:statonary-FP}
\end{eqnarray}
where we wrote the sum over $i$ explicitly.  
Now, considering that
$\partial_\alpha P_{\mathrm{eq}} = -\frac{P_{\mathrm{eq}}}{k_B T} \; \partial_\alpha \mathcal{H}$ and assuming that $P=P_{\mathrm{eq}}$, it follows that the translational terms in Eq.~(\ref{eq:statonary-FP}) are
\begin{eqnarray}
 0 &=&
- \frac{P}{k_B T} \sum_{i\neq j} \frac{p_a^i}{m^i} \, \left[ \frac{\partial J(r_{ij})}{\partial r^i_a} \; \bm{s}_i \cdot \bm{s}_j -  \frac{\partial V(r_{ij})}{\partial r^i_a} \right] 
+
\frac{P}{k_B T} 
\sum_{i \neq j} \left[-\frac{\partial V(r_{ij})}{\partial r^{i}_a}  + \frac{\partial J(r_{ij})}{\partial r^{i}_a} \bm{s}_i \cdot \bm{s}_j \right]\, \frac{p^i_a}{m^i} 
\nonumber\\
&&
+ 3 P \sum_i^N\frac{\gamma_L}{m^i}\, - \frac{\gamma_L P}{k_B T} \sum^N_i \left( \frac{p^i_a}{m^i}\right)^2
+ \frac{D_L P}{k_B T}  \left[
-\sum^N_i\frac{3}{m^i}  + \frac{1}{k_B T} \sum^N_i \left( \frac{p^i_a}{m^i}\right)^2 \; \right] \; . 
\end{eqnarray}
The first line automatically cancels out while the second line leads to the well-known dynamical fluctuation-dissipation relation 
\begin{equation}
D_L = \gamma_L k_B T \; .
\label{dl-tranch}
\end{equation}
This is the exact relationship as derived by Einstein which perfectly matches Eq.~(\ref{DL}).

On the other hand, for the magnetic degrees of freedom and after trivial cancellations of terms we obtain
\begin{eqnarray}
 0 &=& 2 \, 
 \frac{\lambda_s}{s_i} \, \frac{1}{(1+\lambda_s^2)}\sum_i^N  s^i_b \; \omega_b^{i} \; P
- \sum_i^N g^i_{ab} \; \omega_b^{i} \; \frac{\partial P}{\partial s^i_a} 
\nonumber\\
&&
+D_s \frac{1+(A\lambda_s)^2}{(1+\lambda_s^2)^2}\left[-2 \sum_i^N s_a^i \frac{\partial P}{\partial s_a^i} + 
\sum_i^N \left(s_i^2 \delta_{ab} - s_a^i s_b^i\right)\frac{\partial^2 P}{\partial s_a^i \partial s_b^i}\right] 
\; .
\end{eqnarray}
Replacing $g_{ab}^i$ by its explicit expression and the derivatives of $P$, 
$\partial P/\partial s_a^i = \beta\hbar \omega_a^i \, P$
and
$\partial^2 P/\partial s_a^i\partial s_b^j  = (\beta\hbar)^2 \omega_a^i \omega_b^j \,  P$, 
we have that
\begin{eqnarray}
 0 &=& 2 \, \frac{\lambda_s}{s_i} \, 
 \frac{1}{(1+\lambda_s^2)}\sum_i^N  s^i_b \; \omega_b^{i} \; P
- \frac{1}{1+\lambda_s^2} \; \sum_i^N [\epsilon_{abc}s^i_c + \frac{\lambda_s}{s_i} (s_i^2 \delta_{ab} - s^i_a s^i_b) ] \; \omega_b^{i} \; \omega_a^{i} 
\frac{\hbar}{k_B T} \; P
\nonumber\\
&&
+D_s \frac{1+(A\lambda_s)^2}{(1+\lambda_s^2)^2}\left[-2 \sum_i^N s_a^i \omega_a^i \frac{\hbar}{k_B T} \; P+ 
\sum_i^N
\left(s_i^2 \delta_{ab} - s_a^i s_b^i\right) \omega_b^{i} \; \omega_a^{i} \left(\frac{\hbar}{k_B T}\right)^2 P\right] 
\; .
\end{eqnarray}
\end{widetext}
The term proportional to $\epsilon_{abc}$ vanishes because $\epsilon_{abc} \; \omega^i_a \; \omega^i_b \; s^i_c= \bm{\omega}_i \cdot (\bm{\omega}_i \times \bm{s}_i)=0$.  Instead, the remaining terms cancel if we choose
\begin{equation}
D_s =  \frac{\lambda_s(1+\lambda_s^2)}{\left[1+(A\lambda_s)^2 \right]} \frac{k_B T}{\hbar} \; ,
\label{eq:Ds}
\end{equation}

we have already set $s_i=1$. This is the fluctuation-dissipation relation for the magnetic degrees of freedom. Note that for $A=0$ we 
obtain Eq.~(\ref{DS}).  However, it is important to highlight that 
Eq.~(\ref{eq:Ds}) is different from the one reported in 
Ref.~\cite{tranchida2018massively}, which is valid only for damping 
values $\lambda_s \ll 1$.
We have checked, with simulations not shown here, that this is the correct expression that 
allows one to use large values of the damping constant $\lambda_s$ and let the magnetic degrees
freedom equilibrate with the environment.


\bibliography{mainV2}

\newpage

\section*{Supplementary Material}
\beginsupplement

This supplementary material presents extra figures intended to expand the discussion and provide extra evidence and support the points discussed in the main text.

\setcounter{subsection}{0}
\subsection{Average curve}

\begin{figure}[htbp]
\includegraphics[width=0.99\columnwidth]{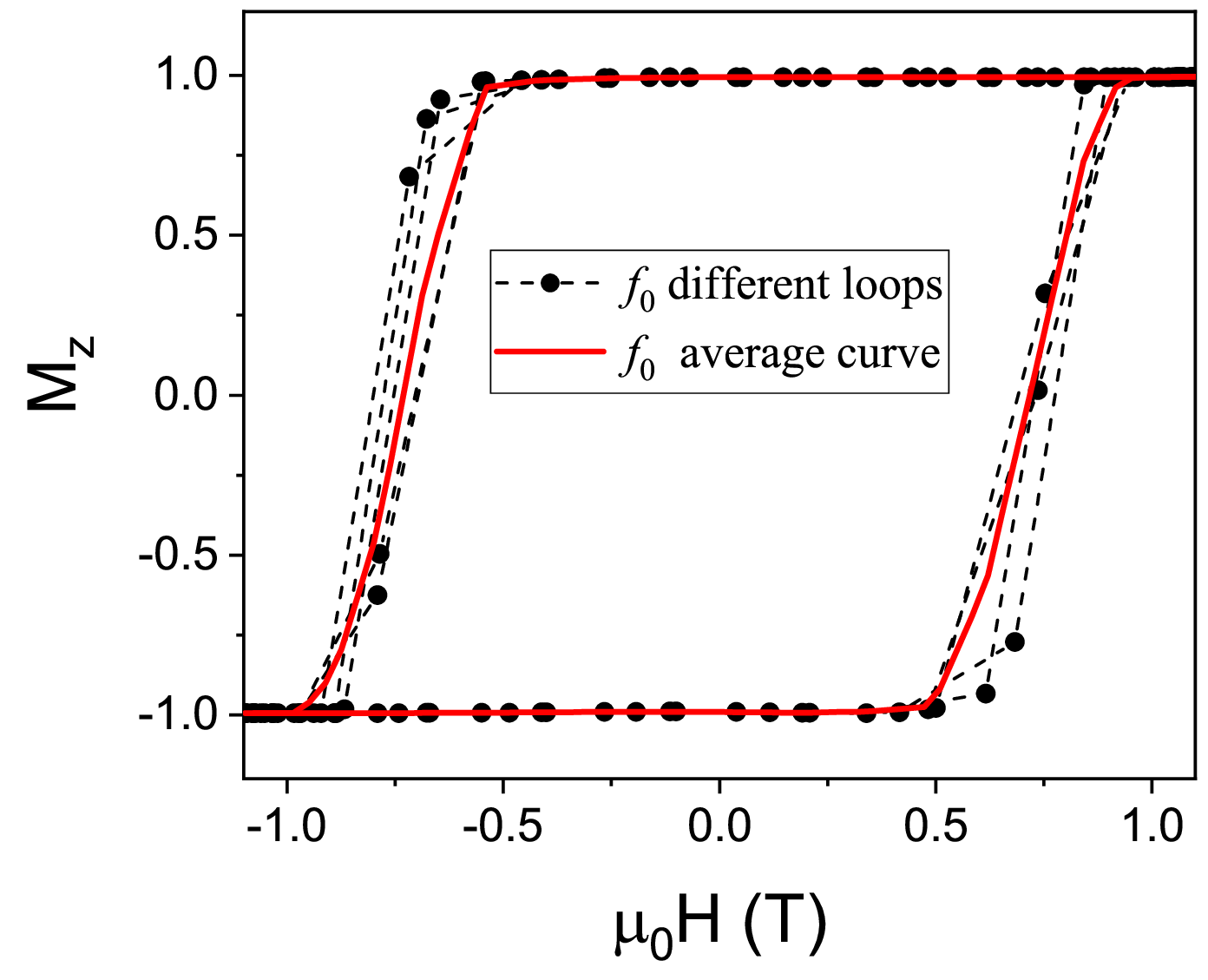}
  \caption{(Color online.) Four different hysteresis loops for the $\phi=0^{\circ}$ case at frequency $f=f_0$ plotted together with the average curve (solid red line).} \label{fig:averagedCycle}
\end{figure}

Figure \ref{fig:averagedCycle} presents an example of how the resulting hysteresis loops where obtained. The resulting curve was obtained by averaging several individual cycles.

\subsection{Effect of frequency and loop convergence}

\begin{figure}[htbp]
\includegraphics[width=0.99\columnwidth]{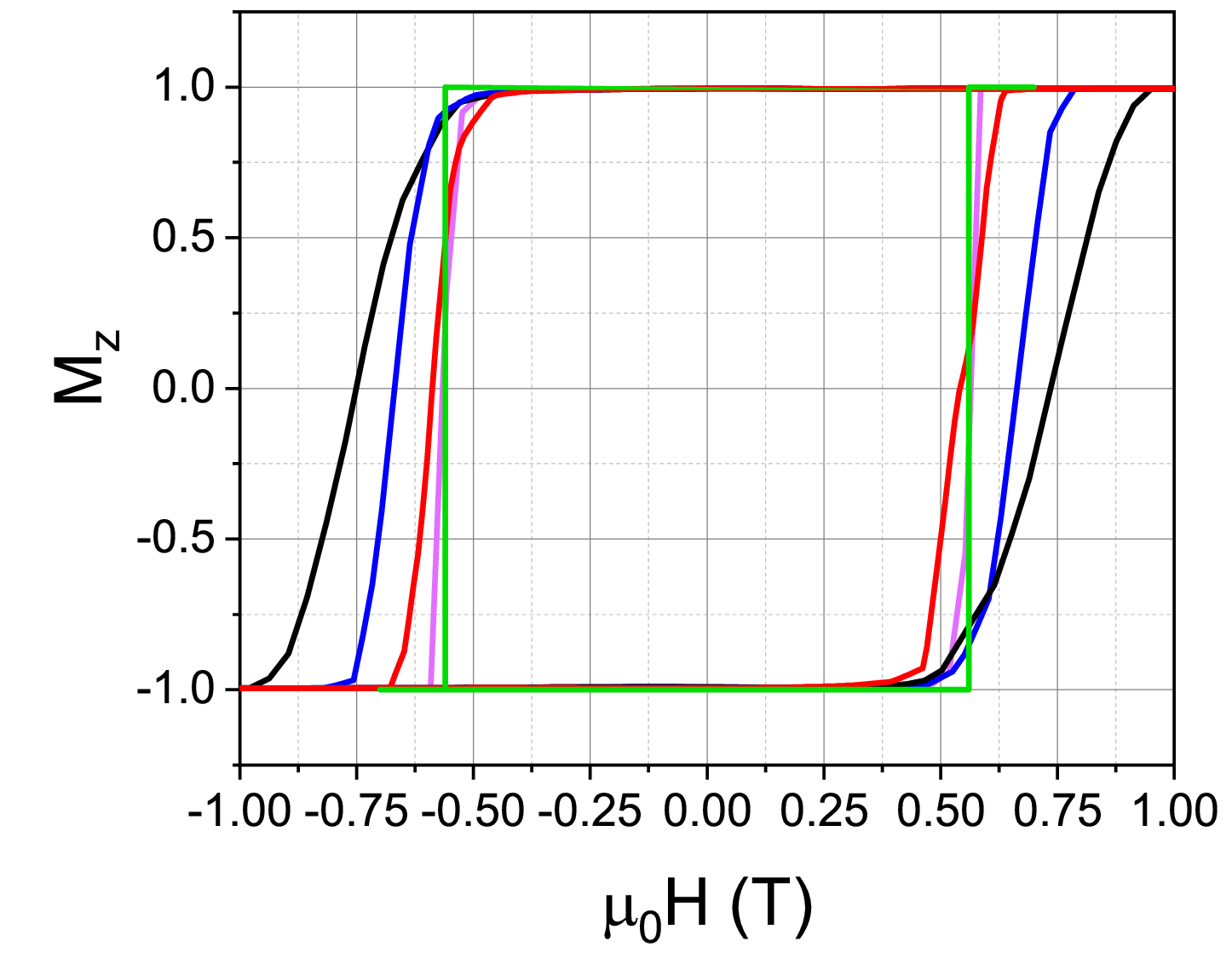}
  \caption{(Color online.) Hysteresis loops for the $\phi=0^{\circ}$ case at different field frequencies $f_0$, $f_0/2$, $f_0/4$ and $f_0/8$, compared to the rectangular loop of the SW model (green curve).} \label{fig:loop_convergence}
\end{figure}

Figure \ref{fig:loop_convergence} shows the hysteresis loops at different frequencies for the case 
in which the anisotropy axis and the direction of the external field form an angle of $\phi = 0^{\circ}$. The plot evidences that the loops narrow when the frequency is decreased until the convergence frequency is found and the resulting loop coincides with the Stoner-Wohlfarth prediction (green curve).
 
\subsection{Increased simulation time}

\begin{figure}[htbp]
\includegraphics[width=0.99\columnwidth]{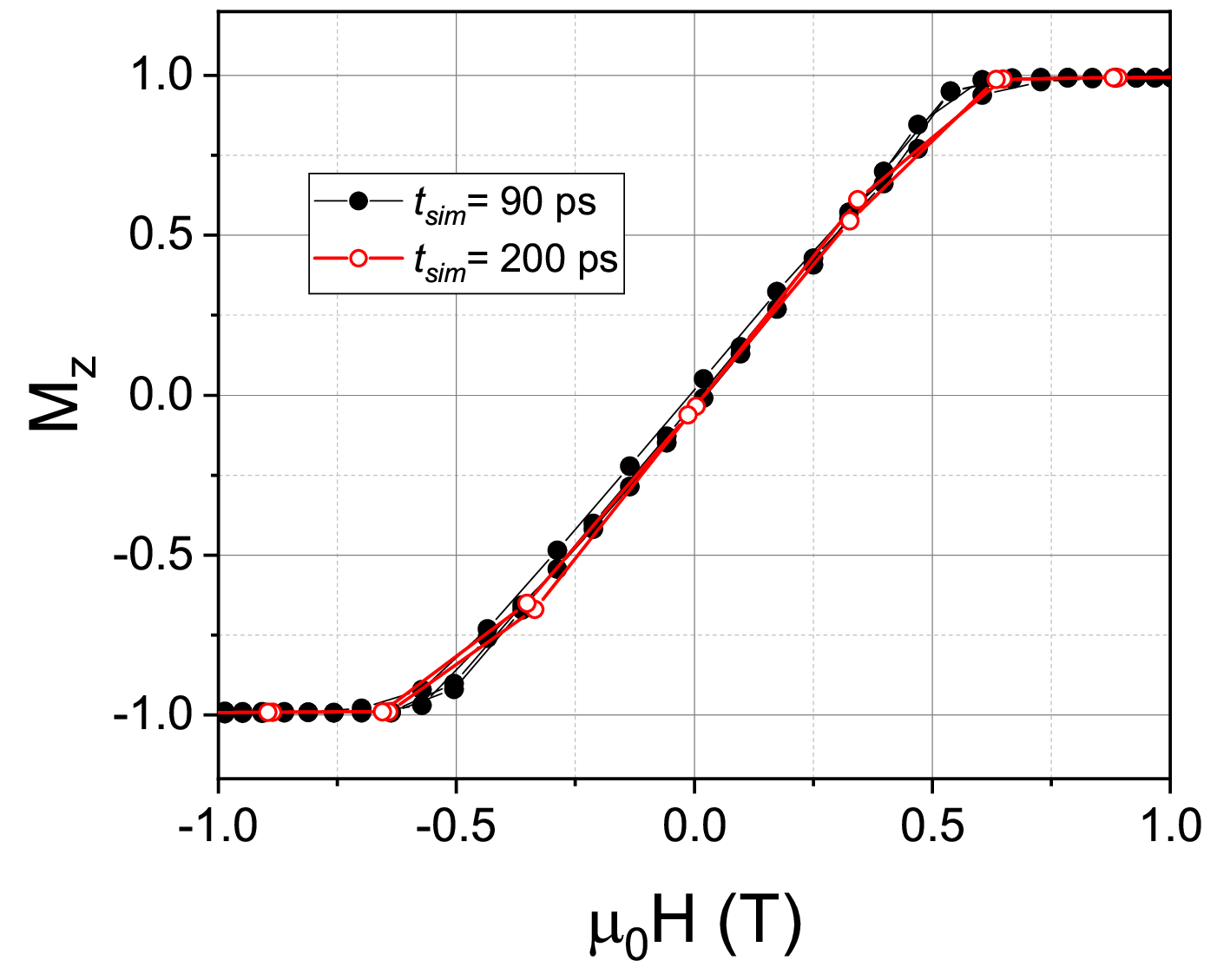}
  \caption{(Color online.) Hysteresis loops for the $\phi=90^{\circ}$ case and frequency $f=f_0/2$ with different simulations times ($t_{\rm sim}$) for each point of the curve} \label{fig:Cycle_vs_time}
\end{figure}

As it is argued in the main text, a simulation time of $t_{\rm sim}=90$ ps for each field value is sufficient to produce reliable hysteresis loops. Figure \ref{fig:Cycle_vs_time} shows that increasing this time by a factor of approximately 2 has no appreciable effect in the shape of the resulting curves.
 
 
 \subsection{Size effects}
 
\begin{figure}[htbp]
\includegraphics[width=0.99\columnwidth]{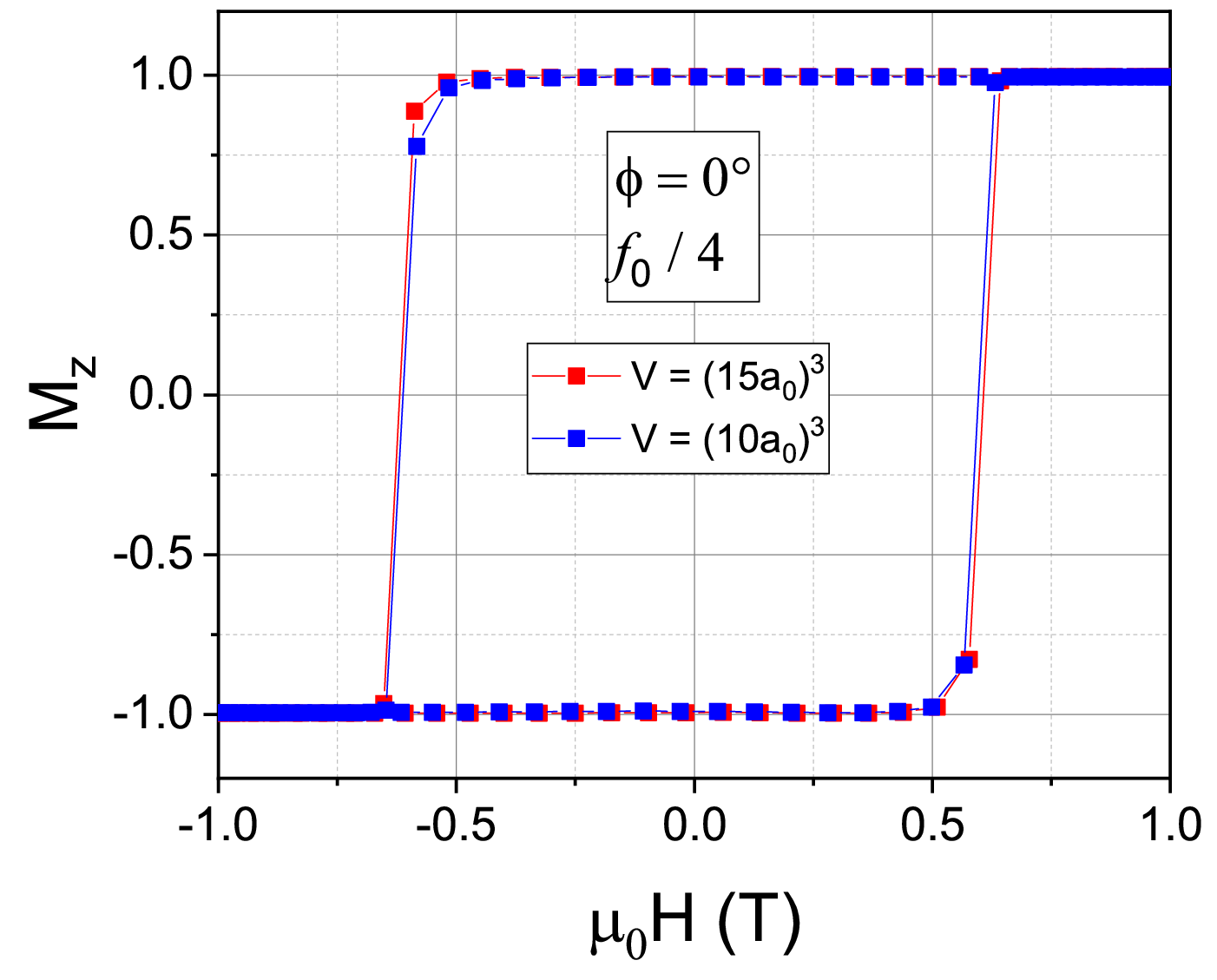}
  \caption{(Color online.) Hysteresis cycles for the $\phi=0^{\circ}$ case and frequency $f_0/4$ for different system sizes.} \label{fig:Cycle_vs_size}
\end{figure}

 Figure \ref{fig:Cycle_vs_size} compares two individual hysteresis loops obtained for a bcc lattice in a cubic box of volume $V=(10 a_0)^3$ (the system under study) and a larger system with $V=(15 a_0)^3$. The simulations correspond to the case of $\phi=0^{\circ}$ and frequency $f_0/4$. 
 
\begin{figure}[htbp]
\includegraphics[width=0.99\columnwidth]{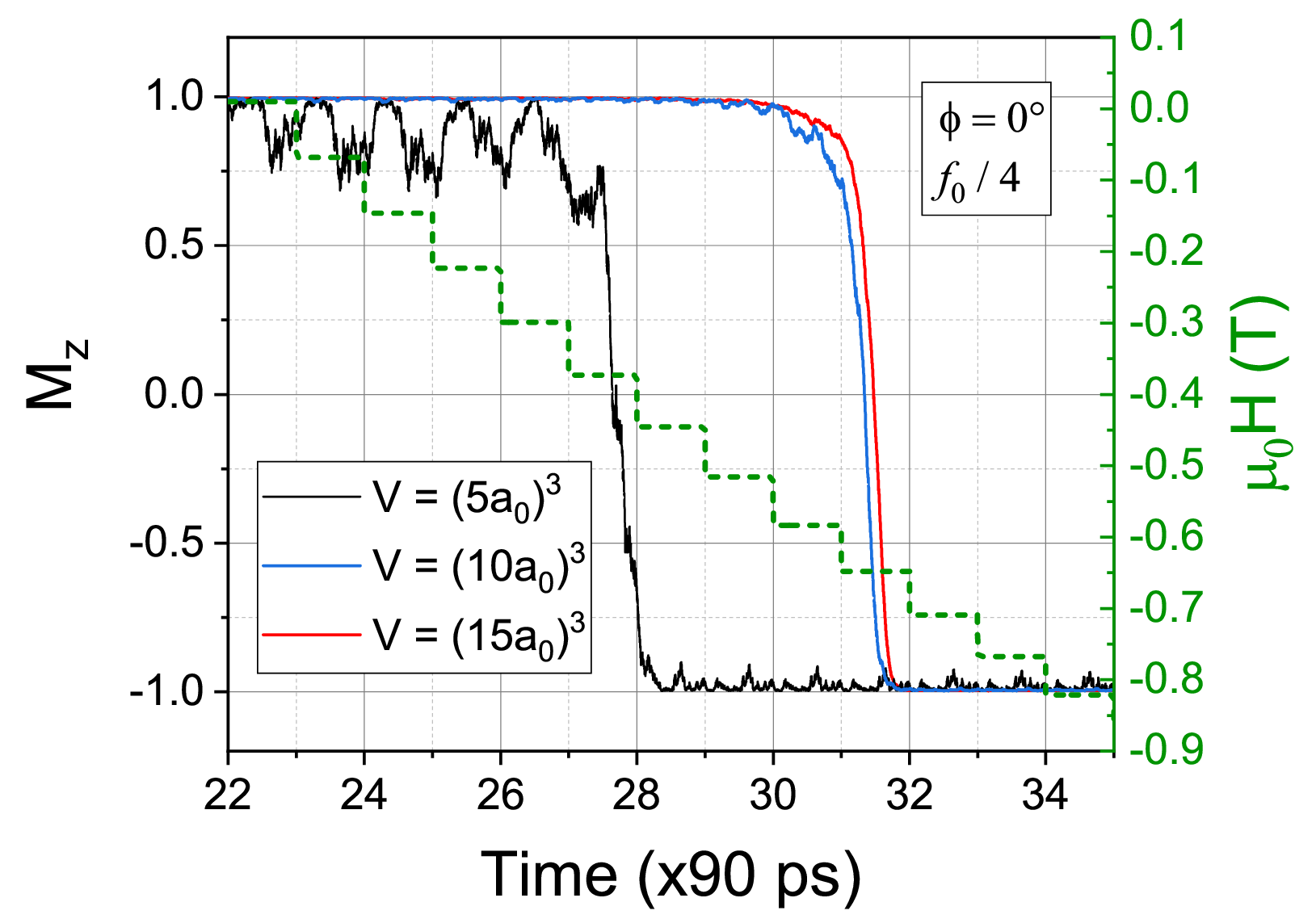}
  \caption{(Color online.) Comparison of the fluctuations of the component of the magnetization along the applied field direction $M_z$ during the time evolution of a simulation throughout half a cycle for different system sizes. The simulations correspond in all cases to the case of $\phi=0^{\circ}$ and frequency $f_0 / 4$. The field value at each time is also included (green dashed line
and right axis).} \label{fig:Equil_vs_size}
\end{figure}

 Figure \ref{fig:Equil_vs_size} compares the fluctuations of $M_z$, the component of the magnetization along the field direction, during  several field steps for three different system volumes. 
 
 Both figures show that simulating a system with $V=(10 a_0)^3$ is sufficient to avoid large size-effects.
 
\subsection{Anisotropy and magnetization fluctuations}

\begin{figure}[htbp]
\includegraphics[width=0.99\columnwidth]{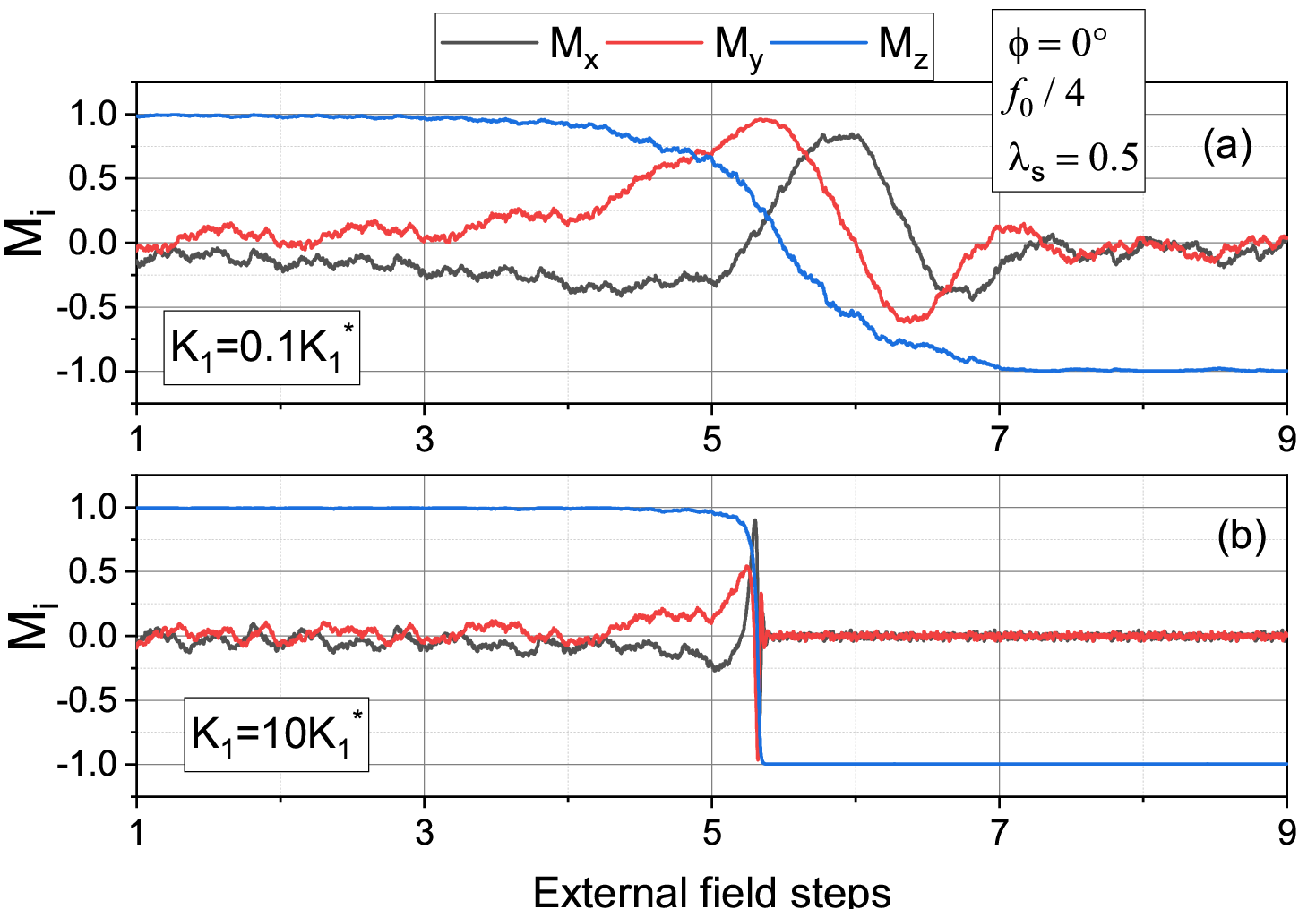}
  \caption{(Color online.) Comparison of the fluctuations of the components of the magnetization for the simulations with $\phi=0^{\circ}$, frequency $f_0 / 4$ and different anisotropy constants, during the same number of applied field steps. The Gilbert damping in these simulations is $\lambda_s=0.5$.} \label{fig:Equil_vs_aniso}
\end{figure}

As it was argued in the Methods section of the main text, we have used mainly uniaxial anisotropy with an anisotropy constant equal to $K_1^*=35$ $\mu$eV/atom, ten times larger than the Fe bulk value. The main reason for this is that, using a large anisotropy in the simulations helps to quickly stabilize the magnetization of the system for the $\phi = 0^{\circ}$ case, and therefore, less simulation time for each field value is required. Figure~\ref{fig:Equil_vs_aniso} shows this effect by comparing the time-evolution of the components of the magnetization for two different anisotropy intensities. 


\subsection{Hysteresis loops and statistics at high temperature}

\begin{figure}[htb]
\includegraphics[width=0.99\columnwidth]{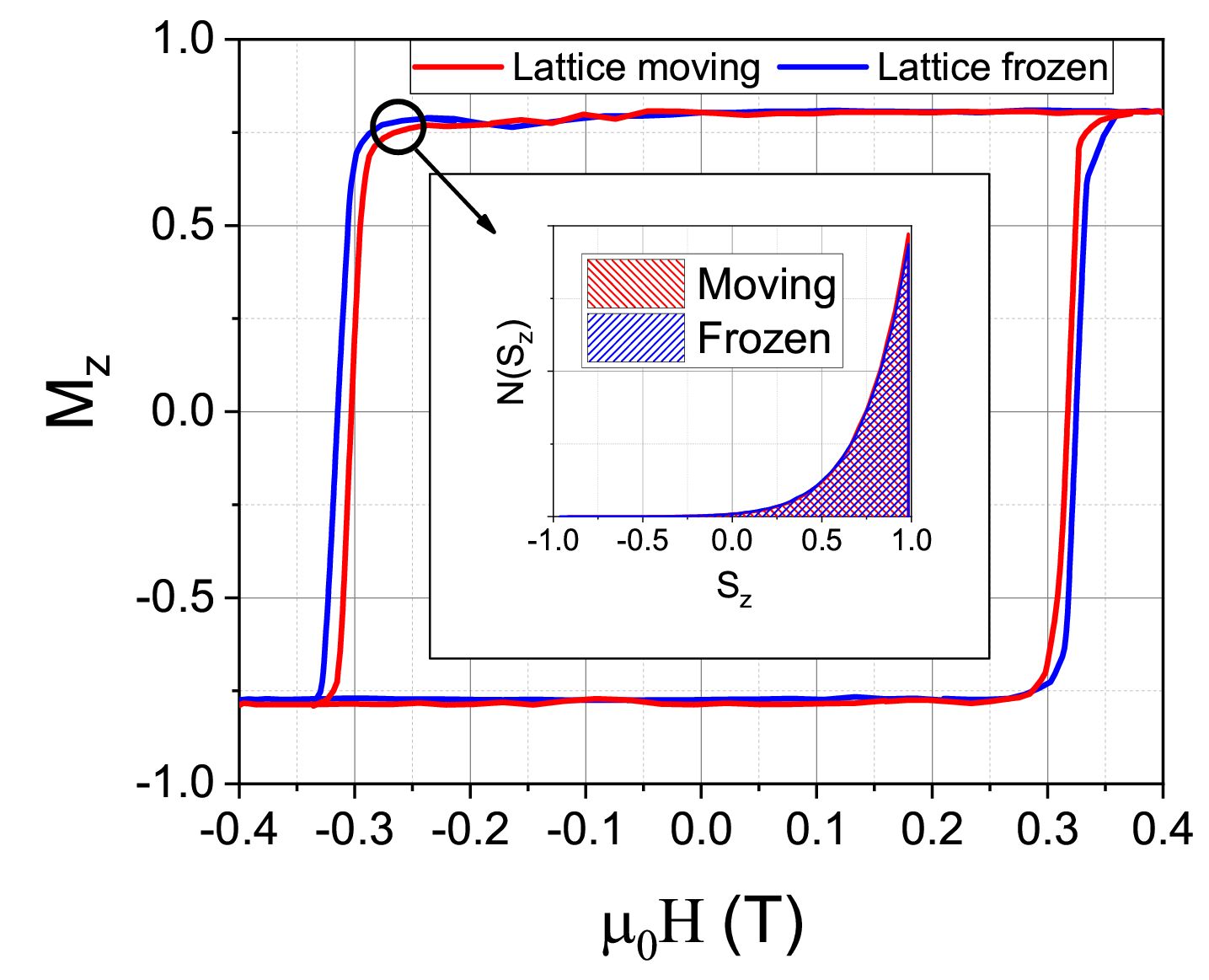}
  \caption{(Color online.) Hysteresis loops obtained at $T=500 K$ under the 2 approaches: lattice frozen and lattice moving. These results correspond to the case of $\phi=0^{\circ}$ and a smaller sweep rate, $SR=0.5 \times 10^8$ $T/s$ (which corresponds to 180ps of simulation time per point). The curves were obtained by averaging over 10 individual loops in each case. The inset shows histograms of the spins orientation along the field direction and corresponds to the state of the system just before the spin flip, indicated by the circle and arrow. These histograms were obtained averaging over 5 different histograms.} \label{fig:lattice_frozen_moving_500K}
\end{figure}

Figure \ref{fig:lattice_frozen_moving_500K} shows the effect of spin-lattice coupling at $T=500$ K. For this case, due to the increased spin fluctuations, the simulations were run for a bcc Fe bulk system with a volume of $(32 \times 32 \times 32)a_0^3$ (65536 atoms), to avoid undesired finite-size effects. In addition, a smaller sweep rate of $SR=0.5 \times 10^8$ $T/s$ (which correspond to 180 ps of simulation time per point) was employed for this simulations. The spin histograms shows that spin fluctuations at this high temperature are comparable for both approaches.

\end{document}